\documentclass[12pt]{article}
\usepackage[utf8]{inputenc}
\usepackage{amsmath,amssymb,aas_macros}
\usepackage{graphicx}
\usepackage{color}
\usepackage{cite}
\usepackage{mathrsfs}

\usepackage[colorlinks,citecolor=blue]{hyperref}
\usepackage{indentfirst}
\makeatletter
\@addtoreset{equation}{section}
\makeatother

\usepackage{comment}
\usepackage{listings}
\begin{document}

\begin{titlepage}

\begin{center}

\vskip .75in

{\Large \bf Constraints on the phase transition of Early Dark Energy
\vspace{2mm}with the CMB anisotropies}

\vskip .75in

{\large
Shintaro Hayashi$\,^{a}$,  Teppei Minoda$\,^{b}$, Kiyotomo Ichiki$\,^{a,c,d}$
}

\vskip 0.25in

{\em
$^{a}$Graduate School of Science, Division of Particle and Astrophysical Science, Nagoya University, Furocho, Chikusa-ku, Nagoya, Aichi 464-8602, Japan
\vspace{2mm} \\
$^{b}$The University of Melbourne, School of Physics, Parkville, VIC 3010, Australia
\vspace{2mm} \\
$^{c}$Kobayashi-Maskawa Institute for the Origin of Particles and the Universe, Nagoya University, Furocho, Chikusa-ku, Nagoya, Aichi 464-8602, Japan
\vspace{2mm} \\
$^{d}$
Institute for Advanced Research, Nagoya University, Furocho, Chikusa-ku, Nagoya, Aichi 464-8602, Japan
}

\begin{abstract}
Early dark energy (EDE) models have attracted attention in the context of the recent problem of the Hubble tension. 
Here we extend these models by taking into account the new density fluctuations generated by the EDE which decays around the recombination phase.
We solve the evolution of the density perturbations in dark energy fluid generated at the phase transition of EDE as isocurvature perturbations. 
Assuming that the isocurvature mode is characterized by a power-law power spectrum and is uncorrelated with the standard adiabatic mode, we calculate the CMB angular power spectra. By comparing them to the Planck data using the Markov-Chain Monte Carlo method, we obtained zero-consistent values of the EDE parameters and $H_0=67.56^{+0.65}_{-0.66}~\mathrm{km} \, \mathrm{s}^{-1} \mathrm{Mpc}^{-1}$ at $68 \%$ CL. This $H_0$ value is almost the same as the Planck value in the $\Lambda$CDM model, $H_0=67.36 \pm 0.54~\mathrm{km} \, \mathrm{s}^{-1} \mathrm{Mpc}^{-1}$, and there is still a $\sim 3.5 \sigma$ tension between the CMB and Type Ia supernovae observations. Including  CMB lensing, BAO, supernovae and SH0ES data sets, we find $H_0=68.94^{+0.47}_{-0.57}~\mathrm{km} \, \mathrm{s}^{-1} \mathrm{Mpc}^{-1}$ at $68 \%$ CL. 
The amplitude of the fluctuations induced by the phase transition of the EDE is constrained to be less than $1$--$2$ percent of the amplitude of the adiabatic mode. This is so small that such non-standard fluctuations cannot appear in the CMB angular spectra.
In conclusion, the isocurvature fluctuations induced by our simplest EDE phase transition model do not explain the Hubble tension well.

\end{abstract}

\end{center}
\vskip .5in

\end{titlepage}

\section{Introduction}
\label{sec:intro}
The Hubble constant $H_0$ is estimated by various observations~\cite{Planck:2018vyg, Riess:2021jrx,2017MNRAS.465.4914B, 2021ApJ...912L...1A, 2021A&A...646A..65M, 2022ApJ...928..165W, 2022JCAP...11..039S}. Recently, a discrepancy arises between values estimated from the early-time and late-time observations.
As an early-time observation, the latest measurements of the cosmic microwave background (CMB) temperature anisotropies by the Planck collaboration give $H_\mathrm{0} = 67.36 \pm 0.54 ~\mathrm{km} \, \mathrm{s}^{-1} \mathrm{Mpc}^{-1}$ by assuming the $\mathrm{\Lambda}$CDM model \cite{Planck:2018vyg}.
On the other hand, the SH0ES Collaboration using Cepheids and Type Ia supernovae shows a higher value as $H_\mathrm{0} = 73.04 \pm 1.04~ \mathrm{km} \, \mathrm{s}^{-1} \mathrm{Mpc}^{-1}$, which are calibrated by Pantheon+ sample
\cite{Riess:2021jrx}.
In addition to the above two examples, other observations estimate $H_0$ (for example, the time delay due to the strong gravitational lensing \cite{2017MNRAS.465.4914B}, the standard sirens from gravitational wave sources \cite{2021A&A...646A..65M} and the age-redshift distribution of the old astrophysical objects \cite{2022ApJ...928..165W} for the late type measurements, CMB polarization anisotropies \cite{2021ApJ...912L...1A} and BAO and BBN data \cite{2022JCAP...11..039S} for the early type measurements).
The difference in estimated $H_\mathrm{0}$ is called "Hubble tension" and some alternative theories beyond $\mathrm{\Lambda}$CDM have been proposed to resolve the Hubble tension
(including, for example, non-cold dark matter \cite{2021arXiv210401077E}, dynamical dark energy \cite{2020PhRvD.102b3518V}, time-varying electron mass \cite{2021PhRvD.103h3507S}, small-scale density fluctuations \cite{2020PhRvL.125r1302J, 2021PhRvD.104j3517R}, varying gravitational constant \cite{2020JCAP...11..024B}, and modified gravity \cite{2020JCAP...10..044B,2021PhRvD.103b3530A,2021PhRvD.103d3528B}).
Another interesting approach is, for example, to estimate the redshift evolution of the Hubble parameter in a phenomenological way, based on the Pantheon sample and BAOs \cite{2021ApJ...912..150D,2022Galax..10...24D}. Their results give a hint to astrophysical systematics or alternative theories.

Early dark energy (EDE) is one of the ideas to solve the Hubble tension \cite{2016PhRvD..94j3523K,2018JCAP...09..025M,2019PhRvL.122v1301P,2021PhRvD.104h3533F,2022arXiv220805583M,2022arXiv220807631R}.
It is motivated by the string-axion, and behaves like the cosmological constant before the critical epoch $a<a_\mathrm{c}$. After that, EDE starts to decay rapidly.
This behavior plays an important role in the estimate of $H_0$ from measurements of the CMB anisotropies.
CMB observations precisely determine the angular size of the sound horizon $\theta_\mathrm{s}^{*}$, which is the ratio between the sound horizon at the last scattering surface $r_\mathrm{s}^{*}$ and the comoving angular diameter distance to the last scattering surface $d_\mathrm{A}^{*}$.
Here, the cosmological constant-like behavior of EDE enhances the expansion rate of the universe before the recombination epoch, and it reduces the sound horizon $r_\mathrm{s}^*$.
Because $\theta_\mathrm{s}^{*}$ must be kept to the observed value, decreasing $r_\mathrm{s}^{*}$ requires the angular diameter distance to be smaller and as a result, $H_0$ is estimated to be larger.

So far, many studies attempted to relieve the Hubble tension by using this enhancement of the expansion rate at early times with an EDE component \cite{2016PhRvD..94j3523K, 2019PhRvL.122v1301P, 2020PhRvL.124p1301S}. They considered the impact of the EDE on the background evolution.
As far as the EDE behaves as the cosmological constant, it never generates the evolution of density perturbations. On the other hand, the density perturbations of EDE are expected to be generated through the gravitational interaction after the EDE starts to decay.
Such EDE perturbations and their important role in the CMB constraints have already been studied and discussed intensively \cite{2019arXiv190401016A,2020PhRvD.102f3527N}.
In addition to those perturbations, because the decay of EDE occurs stochastically at each horizon patch, the isocurvature perturbations could be generated by the phase transition of EDE on the analogy of the bubble nucleation due to the first-order phase transition \cite{1980PhRvD..21.3305C,1983NuPhB.212..321G,1992PhRvD..46.2384T}. This is what we aim to study in this work. In Refs. \cite{2020PhRvD.102f3527N} and \cite{2021PhRvD.104h3533F}, 
the authors argued that such perturbations do not appear on the observational scale of the CMB due to "trigger dynamics" or multiple phase transitions.
In this paper, we treat these isocurvature fluctuations phenomenologically in cosmological perturbation theory and test whether such fluctuations are actually not allowed to exist from CMB observations.

In this paper, we study the effect of the EDE on the CMB anisotropies including the effect of EDE perturbations.
In Sec.~\ref{sec_2}, we introduce the impact of the EDE that has a phase transition before the recombination epoch on the background and perturbation evolutions.
We perform a Markov Chain Monte Carlo (MCMC) analysis to put an observational constraint on the model parameters of the EDE with phase transition and discuss how much the Hubble tension is resolved.
In Sec.~\ref{sec:results}, we show the calculation results for the perturbation equations and the constraints from the MCMC analysis with the Planck CMB measurements. Finally, we discuss and conclude in Sec.~\ref{sec:4}.

\section{Methods}
\label {sec_2}
\subsection{Background effect of early dark energy}
The sound horizon at the last scattering $r_\mathrm{s}^{*}$ and the comoving angular diameter distance to the last scattering surface $d_\mathrm{A}^{*}$ can be written as
    \begin{equation}
    \label{soundhorizon}
        r_\mathrm{s}^{*} = \int_{z*}^{\infty} \frac{dz}{H(z)} c_{s}(z) ,
    \end{equation}
and,
    \begin{equation}
    \label{da}
        d_\mathrm{A}^{*} = \int_{0}^{z^*} \frac{dz}{H(z)}  = \int_{0}^{z^*} \frac{dz}{H_\mathrm{0} \sqrt{\Omega_\mathrm{r0}(1+z)^{4} + \Omega_\mathrm{m0}(1+z)^{3} + \Omega_\mathrm{\Lambda 0}}},
    \end{equation}
respectively, where $z^*$ is the redshift at the last scattering surface and $c_\mathrm{s}(z)$ is the sound speed of the baryon-photon fluid.
Note that we have assumed the standard $\Lambda$CDM model in the second equality in Eq.\eqref{da}.
In the EDE model, one introduces an additional energy component (EDE) in the standard $\Lambda$CDM model, which contributes to the energy density of the universe before recombination. Therefore the EDE reduces $r_\mathrm{s}^{*}$ by increasing $H(z)$ when the $\Lambda$CDM parameters remain unchanged. Recent precise measurements of the CMB anisotropies have determined the angular size of the sound horizon,
\begin{equation}
        \theta_\mathrm{s}^{*} = \frac{r_\mathrm{s}^{*}}{d_\mathrm{A}^{*}}  .
\end{equation}
Since $\theta_\mathrm{s}^{*}$ is tightly constrained by the Planck 2018 CMB data, $\theta_\mathrm{s}^{*}$ must be kept, and therefore, decreasing $r_\mathrm{s}^{*}$ should make $d_\mathrm{A}^{*}$ smaller and $H_0$ larger. 

Our EDE model is based on the phenomenological treatment of an axion-like field, which was introduced in Ref. \cite{2016PhRvD..94j3523K}. In this model, the EDE is assumed to be a slow-roll scalar field in the early epoch and becomes free after the critical epoch $a>a_\mathrm{c}$. Moreover, we assume the energy density of the EDE is converted to the dark radiation (DR) at $a>a_{\rm{c}}$ and the equation of state parameter of DR to be $w_{\rm{DR}}=1/2$ corresponding to the EDE model with the potential $V(\phi) = (1-\mathrm{cos} (\phi/f))^n$ with $n=3$\cite{2020PhRvD.102d3507H}. We simply assume the equation of state of EDE with the above behaviors as

\begin{equation}
w_\mathrm{EDE}=
\begin{cases}
-1 \qquad a < a_\mathrm{c}\\
\phantom{-}\frac{1}{2} \qquad a \geq a_\mathrm{c}
\end{cases}
.
\end{equation}
The total energy density is the sum of the energy densities of the components in the standard model and EDE, and given by 
\begin{align}
\rho_\mathrm{tot} &= \rho_\mathrm{\Lambda CDM} + \rho_\mathrm{EDE} 
\, ,\\
\rho_\mathrm{EDE} &= \rho_\mathrm{EDE}(a_\mathrm{c}) \left(\frac{a_\mathrm{c}}{a} \right)^{3(1+w_\mathrm{EDE})}
.
\end{align}
We introduce $f_\mathrm{EDE}$ which represents a ratio of energy density of EDE to the sum of those of photons, CDM and baryons, and $\rho_\mathrm{EDE}$ is defined as,
\begin{equation}
    \rho_\mathrm{EDE}(a_\mathrm{c}) = f_\mathrm{EDE} \left[ \rho_\gamma (a_\mathrm{c})+ \rho_\mathrm{c} (a_\mathrm{c})+ \rho_\mathrm{c} (a_\mathrm{c})  \right] \, .
\end{equation}
Note that some recent studies have used models with a more realistic potential of the axion fields
\cite{2020PhRvD.101f3523S}, and our EDE model is the basic and simplest one.

\begin{figure}[ht]
    \centering
    \includegraphics{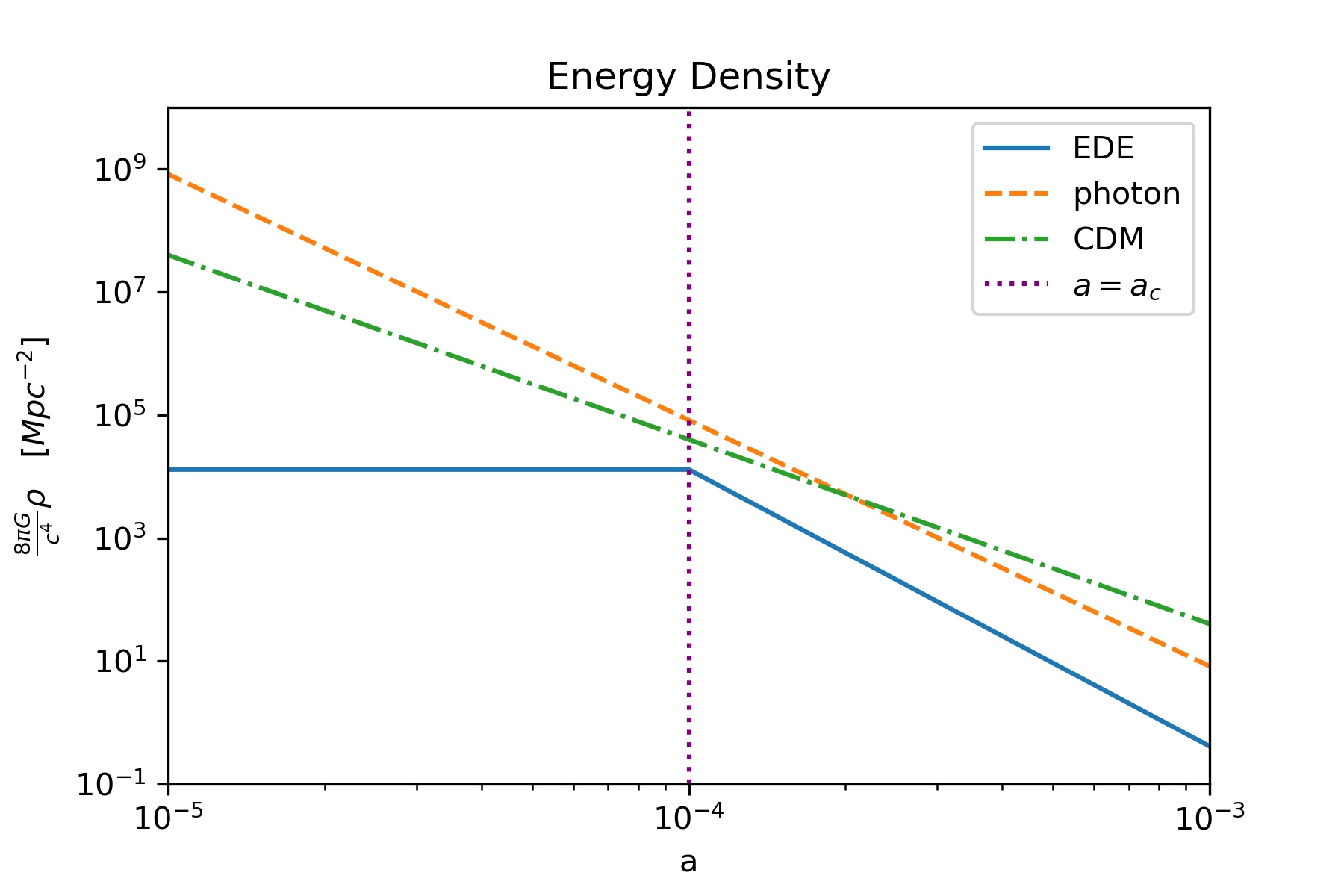}
    \caption{Time evolution of background energy densities of early dark energy ($\rho_{\rm EDE}$; blue solid line), photons ($\rho_\gamma$; orange dashed line), and cold dark matter ($\rho_{\rm CDM}$; green dot-dashed line) as a function of scale factor $a$. For this particular plot, we choose $f_\mathrm{EDE}=0.1$ and $a_\mathrm{c}=10^{-4}$, where $a_c$ is the scale factor when the phase transition occurred.}
    \label{fig_energy_density}
\end{figure}

This is the basic idea of EDE, and there are some concrete models to realize this such as ultra-light axion models \cite{2020PhRvD.101f3523S}.
In Figure \ref{fig_energy_density}, we show the time evolution of energy densities of some components.

\newpage
\subsection{Perturbation evolution of the EDE phase transition mode}

In order to obtain the evolution of perturbations that were excited due to dark energy phase transition, we start at the perturbed energy-momentum conservation equations, $T^{\mu \nu}_{\quad ; \mu} = 0$, which lead to

\begin{equation}
\label{delta}
 \dot{\delta} +(1+ w)\left(\theta + \frac{\dot{h}}{2}\right) +3H \left( \frac{\delta P}{\delta \rho}- w \right) \delta = 0 
 ,
\end{equation}

\begin{equation}
\label{theta}
 \dot{\theta} + H(1-3 w)+\frac{\dot{w}}{1+w} \theta -\frac{\delta P / \delta \rho}{1+ w} k^{2} \delta =0
 ,
\end{equation}
where $\delta$ and $\theta$ are defined as $\delta \equiv \delta \rho / \rho$ and $\theta \equiv i k^{j} v_{j}$.
When treating dark energy as a fluid, one uses a sound speed $c_\mathrm{s}$ which is defined in the dark energy rest frame\cite{2004PhRvD..69h3503B}.
Moreover,  we consider the source of the PT mode perturbation as something similar to an entropy perturbation, and therefore we have
\begin{equation}
\label{source}
    \delta P = c_\mathrm{s}^{2} \delta \rho + 3H(1+w)(c_\mathrm{s}^2-w) \rho \frac{\theta}{k^2} + \rho S
    ,
\end{equation}
where $S$ is the source term which arises during the EDE phase transition ($a_\mathrm{c} \leq a \leq a_\mathrm{end}$).
By combining Eqs.(\ref{delta})-(\ref{source}),
the EDE density perturbation $\delta_\mathrm{EDE}$ and velocity perturbation $\theta_\mathrm{EDE}$ in the synchronous gauge evolve according to the following equations,
\begin{equation}
\label{delta_fin}
\begin{split}
 \dot{\delta}_\mathrm{EDE} +(1+ w_\mathrm{EDE})\left(\theta_\mathrm{EDE} + \frac{\dot{h}}{2}\right) +3H(c_\mathrm{s}^{2}- w_\mathrm{EDE}) \left[\delta_\mathrm{EDE} +3(1+w_\mathrm{EDE})\frac{\theta_\mathrm{EDE}}{k^2}\right]\\ = 
 \left\{
  \begin{array}{ll}
    -3HS     & \qquad (a_\mathrm{c} \leq a \leq a_\mathrm{end} )  \\
    0     & \qquad \mathrm{(others)}
  \end{array}
 \right.
 ,
\end{split} 
\end{equation}

\begin{equation}
\label{theta_fin}
\begin{split}
 \dot{\theta}_\mathrm{EDE} + H(1-3c_\mathrm{s}^{2})\theta_\mathrm{EDE} -\frac{k^2}{1+ w_\mathrm{EDE}}c_\mathrm{s}^{2} \delta_\mathrm{EDE} 
 \\=
 \left\{
  \begin{array}{ll}
   \frac{k^2}{1+ w_\mathrm{EDE}}S     & \qquad (a_\mathrm{c} \leq a \leq a_\mathrm{end} )  \\
    0     & \qquad \mathrm{(others)}
  \end{array}
 \right.
 .
\end{split}
\end{equation}

The perturbations generated by the phase transition (hereafter, PT-mode) may have a $k$-dependence. 
We put the information about $k$-dependence in the initial power spectrum. We assume that the initial power spectrum of PT-mode has a power-law form as
\begin{equation}
    \left<S(\vec{k})S^\ast(\vec{k}')\right>=(2\pi)^3 P_{\rm PT}(k)\delta(\vec{k}-\vec{k}')~,
\end{equation}
where
\begin{equation}
  \frac{k^3}{2\pi^2}  P_\mathrm{PT}(k) =  A_\mathrm{amp}\left( \frac{k}{k_\mathrm{PT}}\right)^{n_\mathrm{PT}-1} 
    .
\end{equation}
Here $A_\mathrm{amp}$, $k_\mathrm{PT}$ and $n_\mathrm{PT}$ are the amplitude, pivot scale and spectral index of the PT-mode, respectively. 
As explained in the following section, $A_\mathrm{amp}$ is replaced with the rescaled parameter $A_\mathrm{PT}\equiv \log_{10} (10^{10} A_\mathrm{amp})$ in our MCMC analysis. Moreover, we use $\mathrm{log}_{10}(a_\mathrm{end}/a_\mathrm{c})$ as the parameter in MCMC to prevent $a_\mathrm{end}$ from becoming smaller than $a_\mathrm{c}$.

\section{Results and Discussion}
\label{sec:results}
\subsection{Evolution of the density perturbation}
\label{re_pertub}
\begin{figure}[ht]
    \centering
    \includegraphics[width=1.0\linewidth]{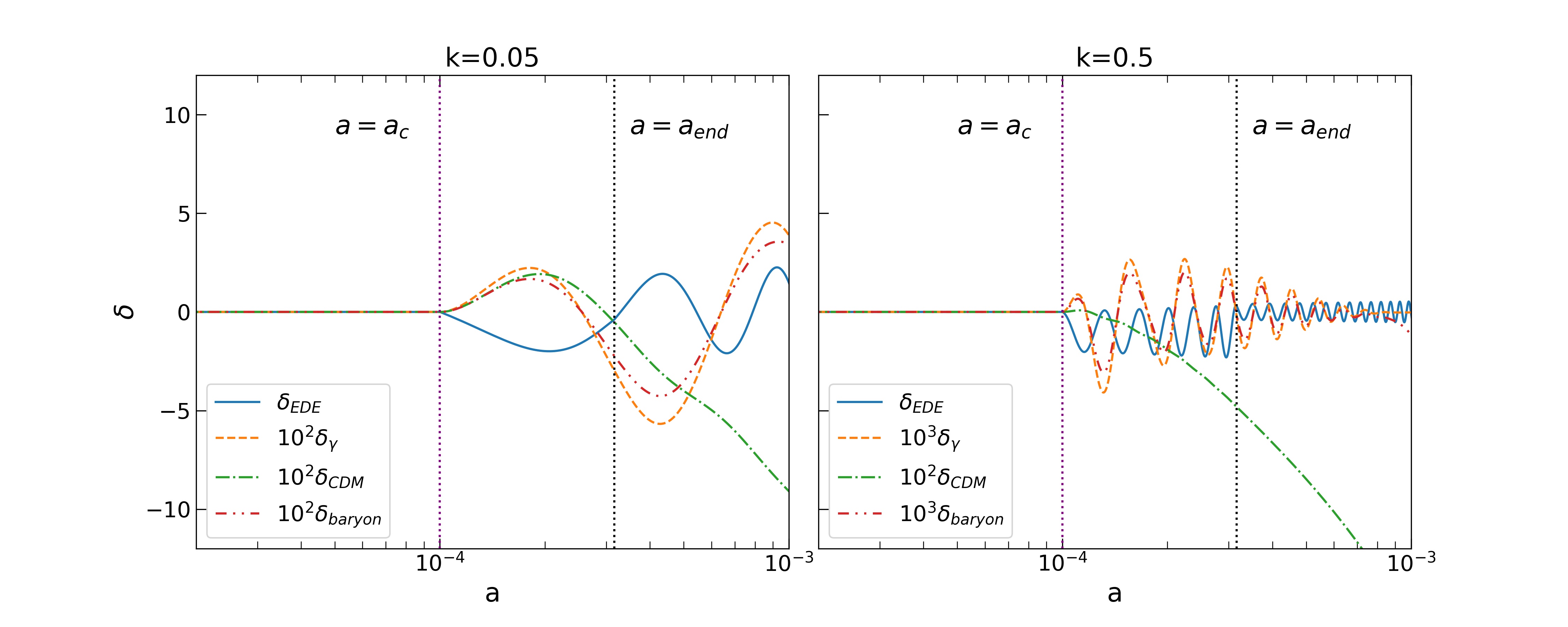}
    \caption{Time evolution of density perturbations of EDE ($\delta_{\rm EDE}$, solid), photons ($\delta_{\gamma}$, dashed), cold dark matter ($\delta_{\rm CDM}$, dot-dashed), and baryons ($\delta_{\rm baryon}$, dotted), at wavenumber $k=0.05$ [Mpc$^{-1}$] (left panel) and $k=0.5$ [Mpc$^{-1}$] (right panel). The purple and black dashed vertical lines show the beginning and end of the PT. The perturbations are generated at $a=a_{\rm c}$ and grow up after that. The oscillation center slightly changes at $a=a_{\rm end}$, respectively. In this figure, the amplitudes of $\delta_{\gamma}$, $\delta_{\rm CDM}$ and $\delta_{\rm baryon}$ are multiplied by a factor of $10^{2}$ or $10^{3}$ for clarity.}
    \label{fig_per}
\end{figure}

We show the results of perturbation evolution of the PT-mode at some wavenumbers in Figure \ref{fig_per}. We assumed the perturbation evolution of the PT-mode is independent of the adiabatic mode, and here we set the initial conditions for the adiabatic mode perturbations as zero, in order to simply illustrate the PT-mode perturbations. Therefore, it can be seen that there is no fluctuation until PT occurs and $\delta_\mathrm{EDE}$ starts to oscillate when the source term appears at $a=a_\mathrm{c}$ when PT happens. In the larger $k$-mode, the faster the EDE density perturbation oscillates. The EDE perturbation gravitationally propagates to other perturbations. 
At first, the oscillation phase of density perturbation of other components is inverse of that of EDE. This is because the local expansion rate is faster and the energy densities of the other fluid components are lower where the EDE energy density is higher than the background. 
In the early universe, baryons are tightly coupled with photons, and the density perturbations of the baryon and photon before recombination evolve together. The CDM density perturbation is initially generated by that of the EDE gravitationally. Once it is generated, it grows up by its self-gravity.

The end of the PT period is shown as the vertical black line in Figure \ref{fig_per}.
Although we can see a small effect on the evolution of density perturbations from the end of the PT in the change of the oscillation center, it is clear that the gravitational growth and the acoustic oscillation of the perturbation dominate after PT happens.

\subsection{Power spectra of CMB anisotropies from the phase transition mode}
\label{sec:results_CMB}
\begin{figure}[ht]
    \centering
    \includegraphics[width=15cm]{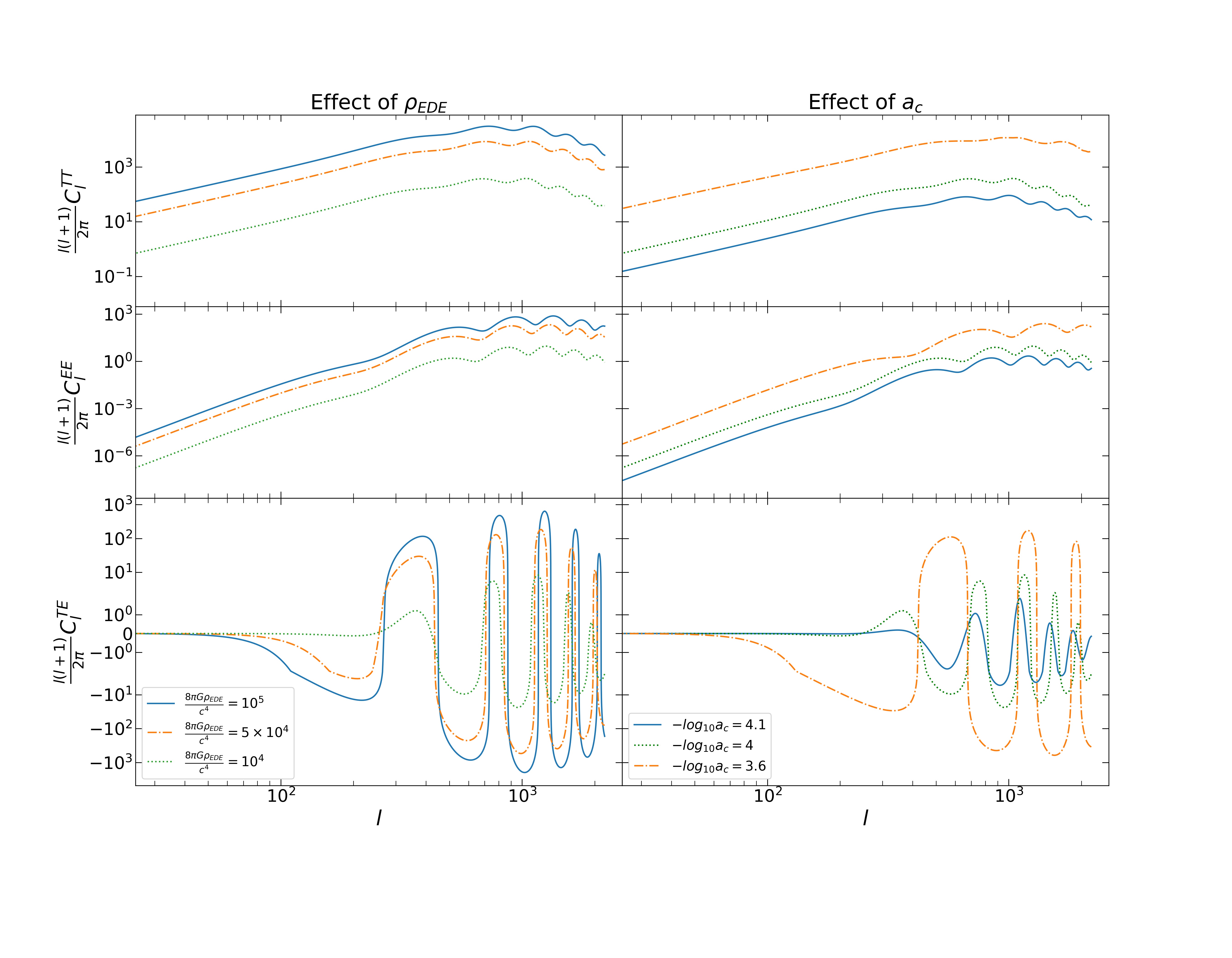}
    \caption{The CMB angular power spectra of the PT-mode.  We only show the multipole range of $l \geq 25$ because the spectra simply decay with a power-law at $l \leq 25$. In the left panel, the initial EDE energy density is taken as $\rho_\mathrm{EDE}$, $1 \times 10^{4}$, $5 \times 10^{4}$ and $1 \times 10^{5}$ with $-\mathrm{log}_{10}a_\mathrm{c}$ fixed to $4$. In the right panel, we set $f_\mathrm{EDE}=0.1$ and vary $a_\mathrm{c}$ as $-\mathrm{log} a_\mathrm{c} = 4.1$, $4$ and $3.6$. In both cases, the other EDE and PT related parameters are fixed to $a_{\rm{end}}=3 \times 10^{-4}$ and $A_{\rm PT}=4$.}
    \label{fig_clfedeac}
\end{figure}

\begin{figure}[ht]
    \centering
    \includegraphics[width=15cm]{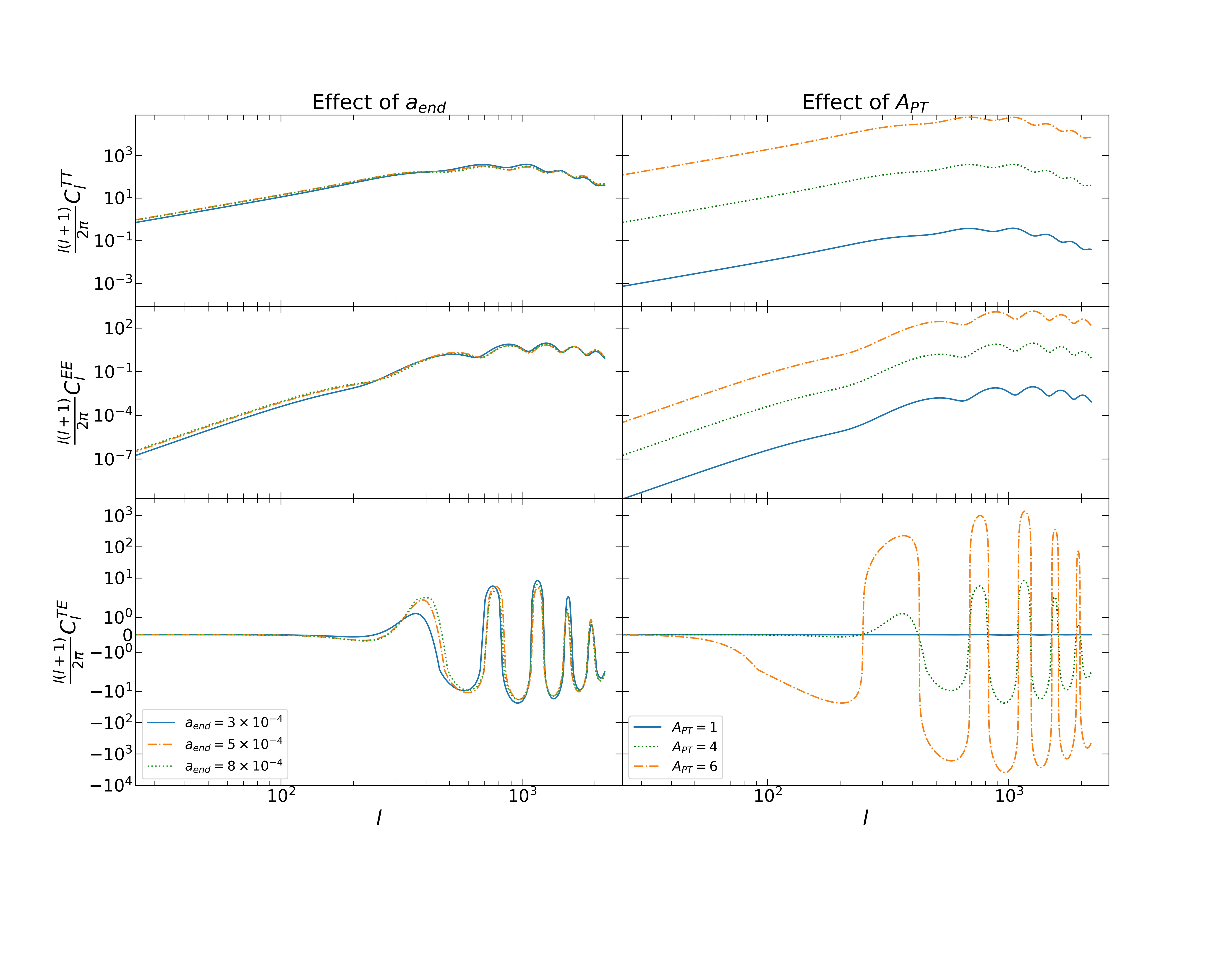}
    \caption{We show the effects of $a_\mathrm{end}$ in the left panel and effects of $A_\mathrm{PT}$ in the right panel. In the left panel, the end of the phase transition is taken as $a_\mathrm{end}$, $3 \times 10^{-4}$, $5 \times 10^{-4}$ and $8 \times 10^{-4}$ with $A_\mathrm{PT}=4$. In the right panel, we set $a_\mathrm{end}=3 \times 10^{-4}$ and vary $A_\mathrm{PT}$ as $A_\mathrm{PT} = 1$, $4$ and $6$. In both cases, the other EDE and PT related parameters are fixed to $\rho_\mathrm{EDE}= 10^{4}$ and $-\mathrm{log}_{10}a_\mathrm{c}=4$.}
    \label{fig_claendapt}
\end{figure}

\begin{figure}[ht]
    \centering
    \includegraphics[width=15cm]{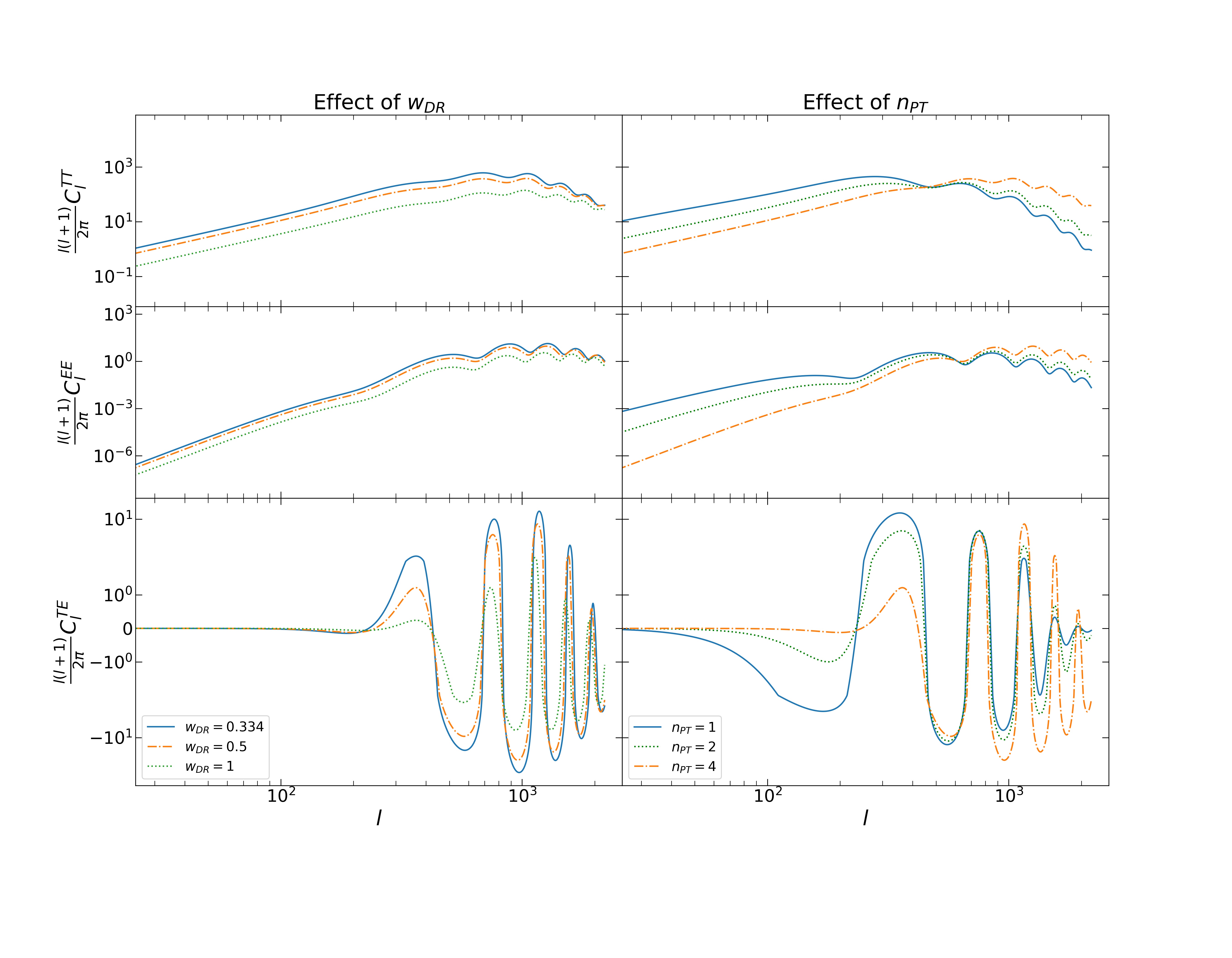}
    \caption{We show the effects of $w_\mathrm{DR}$ in the left panel and effects of $n_\mathrm{PT}$ in the right panel. In the left panel, the end of the phase transition is taken as $w_\mathrm{DR}$, $0.334$, $0.5$ and $1$ with $n_\mathrm{PT}=4$. In the right panel, we set $w_\mathrm{DR}=0.5$ and vary $n_\mathrm{PT}$ as $n_\mathrm{PT} = 1$, $2$ and $4$. In both cases, the other EDE and PT related parameters are fixed to $\rho_\mathrm{EDE}= 10^{4}$, $-\mathrm{log}_{10}a_\mathrm{c}=4$, $a_\mathrm{end}=3 \times 10^{-4}$ and $A_\mathrm{PT}=4$.}
    \label{fig_clwdrnpt}
\end{figure}

First, we show the effect of the PT-mode perturbations on the CMB angular power spectrum $C_\ell^{\rm{PT}}$ in Figure \ref{fig_clfedeac}, \ref{fig_clwdrnpt} and \ref{fig_claendapt}. In the left panel of Figure \ref{fig_clfedeac}, we have chosen the initial energy densities of the EDE as $8 \pi G \rho_{\rm{EDE}} /c^4 = 10^{4}$, $5 \times10^{4}$ and $10^{5}$ [$\rm{Mpc}^{-2}$]. Although we discuss the effects by $\rho_\mathrm{EDE}$ here, we use the $f_\mathrm{EDE}$ as the parameter in our MCMC analysis. It can be seen that $\rho_\mathrm{EDE}$ changes the amplitude of the spectra and the higher energy density of EDE leads to the larger amplitude of spectra. The peak positions slightly shift to the smaller scale when $\rho_\mathrm{EDE}$ becomes larger because the sound horizon becomes smaller. In the right panel of Figure \ref{fig_clfedeac}, we also show the angular power spectra with different phase transition times of the EDE $a_\mathrm{c}$. According to this figure, $a_\mathrm{c}$ also changes the amplitudes and the peak positions. The smaller $a_{\rm{c}}$ is, the earlier EDE decays and the lower is the ratio of the EDE energy density to the total energy density during the phase transition.
Therefore, the amplitude of spectra becomes smaller due to the smaller early-ISW effect. The parameter $a_{\rm{c}}$ also controls the beginning of the phase transition, and the smaller $a_{\rm{c}}$ shifts the peak positions to the larger scale.

In Figure \ref{fig_claendapt}, we show that how $a_\mathrm{end}$ and $A_\mathrm{PT}$ affect the $C^\mathrm{PT}_\ell$. It is clear that $A_\mathrm{PT}$ only affects the amplitude of the spectra. 
As we saw previously, the growth of the EDE density perturbation is dominated by its self-gravity after the source term appears at $a_{\rm{c}}$. Therefore, the effect of $a_{\rm{end}}$ on the perturbation evolution is small. According to the left panel of Figure \ref{fig_claendapt}, the effect of $a_\mathrm{end}$ is not significant compared to other parameters. Note that these two parameters do not alter the background evolution. On the other hand, $\rho_\mathrm{EDE}$ and $a_\mathrm{c}$ affect the background level as we explained in Section \ref{sec_2}, and thus they also affect the adiabatic mode perturbations.

The dependence of $C_\ell^\mathrm{PT}$ on $w_\mathrm{DR}$ and $n_\mathrm{PT}$ is shown in Figure \ref{fig_clwdrnpt}. The change of $w_\mathrm{DR}$ corresponds to the change of the decay speed of EDE. Therefore, the smaller $w_\mathrm{DR}$ leads to the larger amplitude of spectra on large scales. Note that, however, these parameters are fixed to $w_\mathrm{DR}=1/2$, $n_\mathrm{PT}=4$ in our MCMC analysis.

\begin{figure}[ht]
    \centering
    \includegraphics[width=15cm]{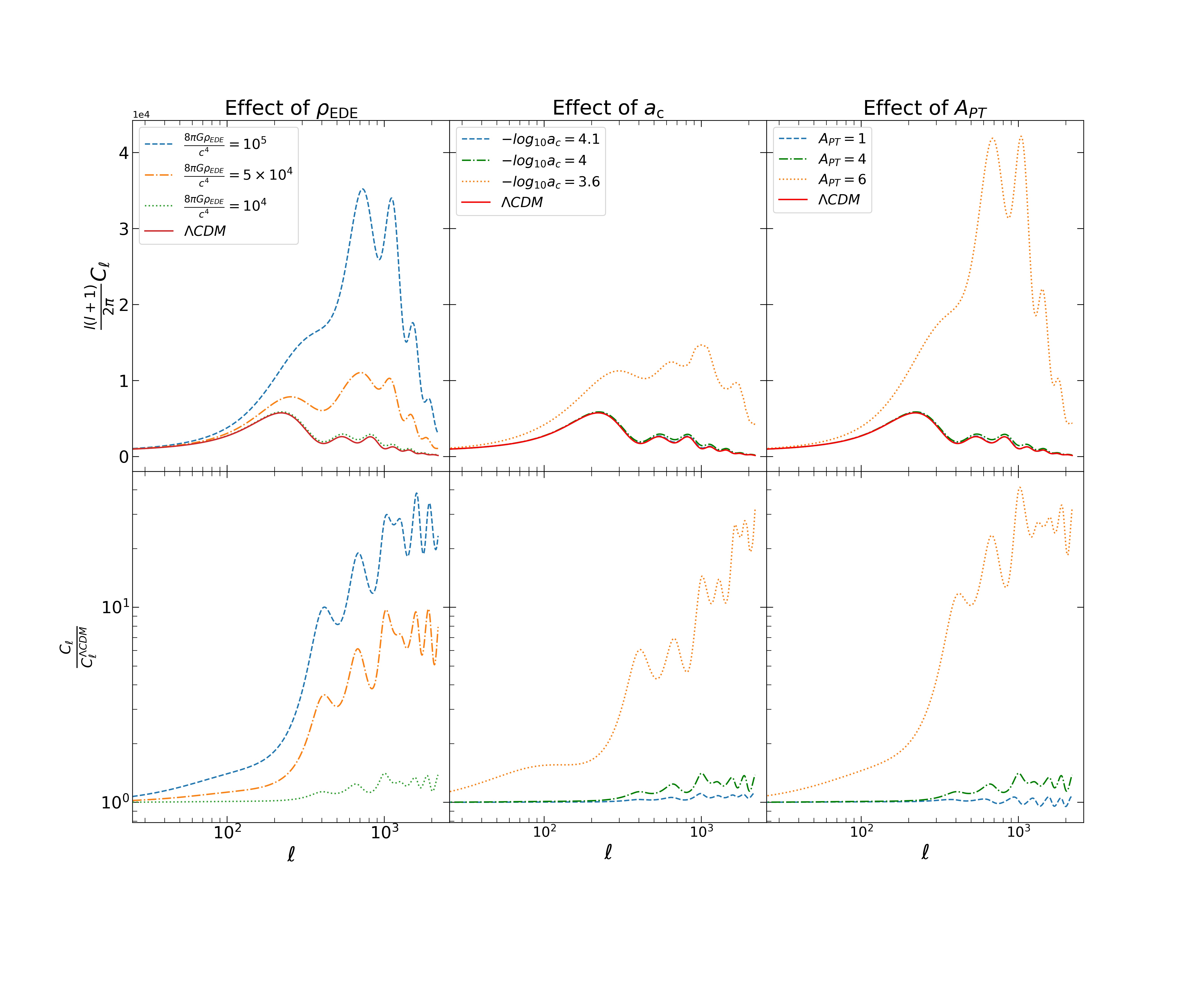}
    \caption{The effects of $\rho_\mathrm{EDE}$, $a_\mathrm{c}$ and $A_\mathrm{PT}$ on the total TT spectra are shown in the upper panels. The red solid line shows the CMB spectra made by $\Lambda$CDM best-fitted parameters. In the lower panels, the fraction of each spectra to the $\Lambda$CDM spectra are shown. Basically, the other EDE parameters that we do not change are fixed to $\rho_\mathrm{EDE}= 10^{4}$, $-\mathrm{log}_{10}a_\mathrm{c}=4$, $a_\mathrm{end}=3 \times 10^{-4}$, $A_\mathrm{PT}=4$, $n_\mathrm{PT}=4$ and $w_\mathrm{DR}=0.5$. }
    \label{fig_cltotad1}
\end{figure}

\begin{figure}[ht]
    \centering
    \includegraphics[width=15cm]{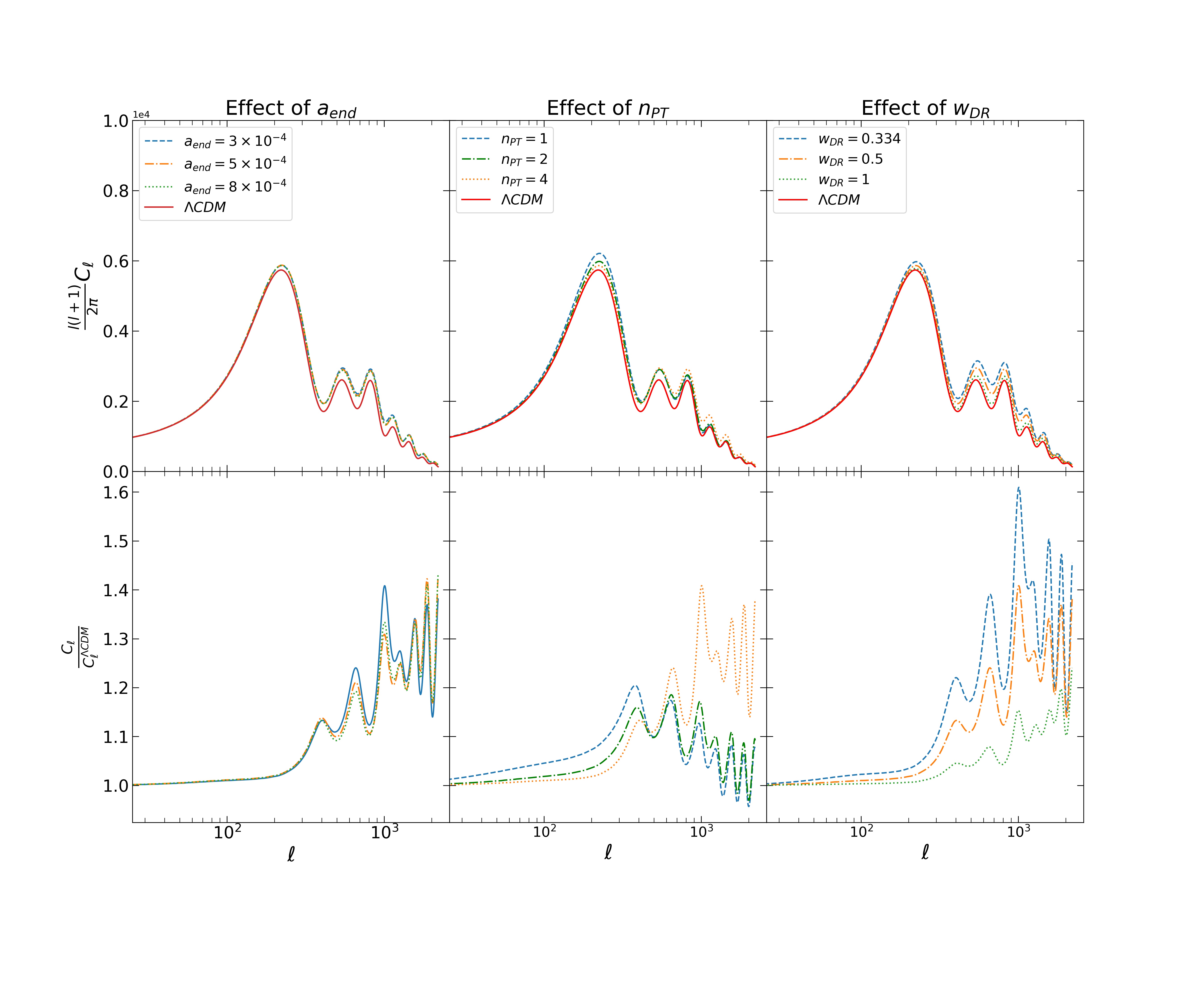}
    \caption{The effects of $a_\mathrm{end}$, $n_\mathrm{PT}$ and $w_\mathrm{DR}$ on the total TT spectra are shown in the upper panels. The red solid line shows the CMB spectra made by $\Lambda$CDM best-fitted parameters. In the lower panels, the fraction of each spectra to the $\Lambda$CDM spectra are shown. Basically, the other EDE parameters that we do not change are fixed to $\rho_\mathrm{EDE}= 10^{4}$, $-\mathrm{log}_{10}a_\mathrm{c}=4$, $a_\mathrm{end}=3 \times 10^{-4}$, $A_\mathrm{PT}=4$, $n_\mathrm{PT}=4$ and $w_\mathrm{DR}=0.5$.}
    \label{fig_cltotad2}
\end{figure}

Next, we show the effects of each parameter on the total CMB TT angular power spectrum in the upper panels of Figure \ref{fig_cltotad1} and \ref{fig_cltotad2}. The red solid lines are the TT spectra generated using the $\Lambda$CDM best-fitted parameters\cite{Planck:2018vyg}. The other lines are the CMB spectra, which contain the EDE and PT-mode perturbations. By including the PT-mode, the CMB spectra are amplified, and in particular, its effects are large at small scales. Furthermore, we plot the fraction of each spectrum to the $\Lambda$CDM one in the lower panels of Figure \ref{fig_cltotad1} and \ref{fig_cltotad2}. The spectra which are substantially different from the red solid lines, such as the blue-dashed line in the left-upper panel in Figure \ref{fig_cltotad1}, are rejected by the CMB data.

\subsection{Implementation into \texttt{CosmoMC}}
For the MCMC analysis, we have used the \texttt{CosmoMC} package \cite{Lewis:2013hha,Lewis:2002ah}. In the analysis of Section~\ref{MCMC_Planck2018}, we use the Planck 2018 low-$\ell$ and high-$\ell$ temperature and polarization data \cite{2020A&A...641A...5P}.
In addition to the Planck data, we also include the CMB lensing \cite{2020A&A...641A...8P}, BAO, supernovae, and local distance ladder in Section~\ref{MCMC_full}. BAO data is constructed by the 6dF Galaxy Survey\cite{2011MNRAS.416.3017B}, SDSS DR7 Main Galaxy Sample\cite{2015MNRAS.449..835R} and the SDSS BOSS DR12\cite{2017MNRAS.470.2617A}. We use the Pantheon data set of type Ia Supernovae\cite{2018ApJ...859..101S} as the supernovae data, and SH0ES results $H_0=73.04 \pm 1.04$ as the local distance ladder data.

We have modified CAMB \cite{Lewis:1999bs} to calculate the perturbation evolution and the CMB spectra made by PT. We have added four new parameters, $- \mathrm{log}_{10}a_\mathrm{c}$, $\mathrm{log}_{10} (a_\mathrm{end} / a_\mathrm{c})$, $f _\mathrm{EDE}$ and $A_\mathrm{PT}$ into CAMB and they are the free parameters in the MCMC in addition to the standard six cosmological parameters. Here, $A_\mathrm{PT}$ is defined as $A_\mathrm{PT} \equiv {\rm{log}}_{10}(10^{10}A_{\rm amp})$ for convenience. Although we have also put $n_{\rm PT}$ into CAMB as a parameter, in the following analysis, it is not the free parameter and is fixed to 4 in order to take into account that the phase transition occurred spatially at random and the perturbations have a white noise spectrum.
We put the flat priors on the parameters and show them in Table \ref{prior_params}. We use the same ranges of the priors in all the analyses except for the second one in Section~\ref{MCMC_Planck2018}.
The MCMC sampling stops
when the convergence reaches the Gelman and Rubin $R$ condition \cite{10.1214/ss/1177011136} and we set $R-1<0.01$ for all our analysis.

\begin{table}[ht]
    \centering
    \renewcommand{\arraystretch}{1.5}
    \scalebox{0.8}{
    \begin{tabular}{|c|c|c|}
    \hline
    & Parameter & Priors \\
    \hline
    $\Lambda$CDM parameters & $\Omega_{\rm b} h^2$ & [0.05, 0.1] \\
    & $\Omega_{\rm c} h^2$ & [0.001, 0.99] \\
    & $\tau$ & [0.01, 0.8] \\
    & $100\Theta_{\rm{s}}$ & [0.5, 10] \\
    & $n_{\rm{s}}$ & [0.7, 1.2] \\
    & $\mathrm{log}_{10}A_{\rm{s}}$ & [2, 5] \\
    \hline
    EDE parameters & $f_\mathrm{EDE}$ & [0, 0.1] \\
    & $- \mathrm{log}_{10}a_\mathrm{c}$ & [3.4, 5] \\
    \hline
    PT parameters & $A_{\rm{PT}}$ & [1.5, 7.5] \\
    & $\mathrm{log}_{10}(a_{\rm{end}}/a_\mathrm{c})$ & [0.1, 1] \\
    \hline
    \end{tabular}
    }
    \caption{The priors distribution on the parameters are shown.}
    \label{prior_params}
\end{table}

\subsection{Constraints on EDE model with MCMC}

When we plot the results of the MCMC analysis, the blue contours show the constraints in the $\Lambda$CDM model, and the green contours marked as ``EDE+PT'' indicate the constraints in the EDE model with the phase transition. We also show the case of the EDE model without PT-mode indicated by the red contour labeled ``EDE'' to compare the results of our EDE model with those of the simple EDE model used in previous studies \cite{2019PhRvL.122v1301P,2022ApJ...929L..16H,2022arXiv220902708C}.

\subsubsection{Planck2018}
\label{MCMC_Planck2018}
\begin{figure}[ht]
    \centering
    \includegraphics[height=16cm]{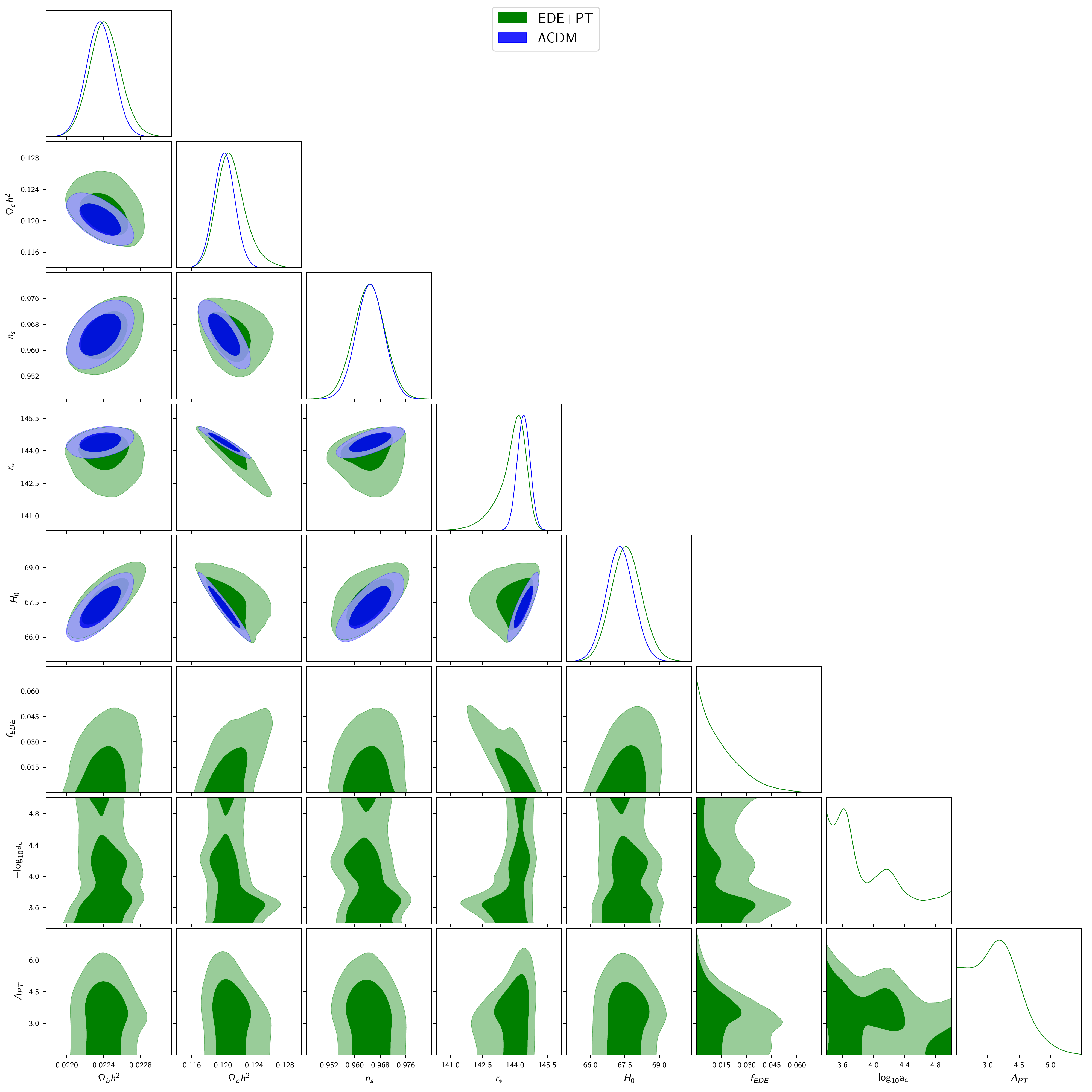}
    \caption{68\% and 95\% confidence level obtained from the MCMC analysis using the Planck 2018 data. The blue and green contours show the $\Lambda$CDM and EDE+PT model.}
    \label{fig_mcmc_3411}
\end{figure}

We consider the Planck 2018 temperature and polarization data in this part. We plot the results of the MCMC analysis in Figure \ref{fig_mcmc_3411} and show the best-fitted values and constraints with 68\% confidence level in Table \ref{tab_params}.

\begin{figure}[ht]
    \centering
    \includegraphics[height=16cm]{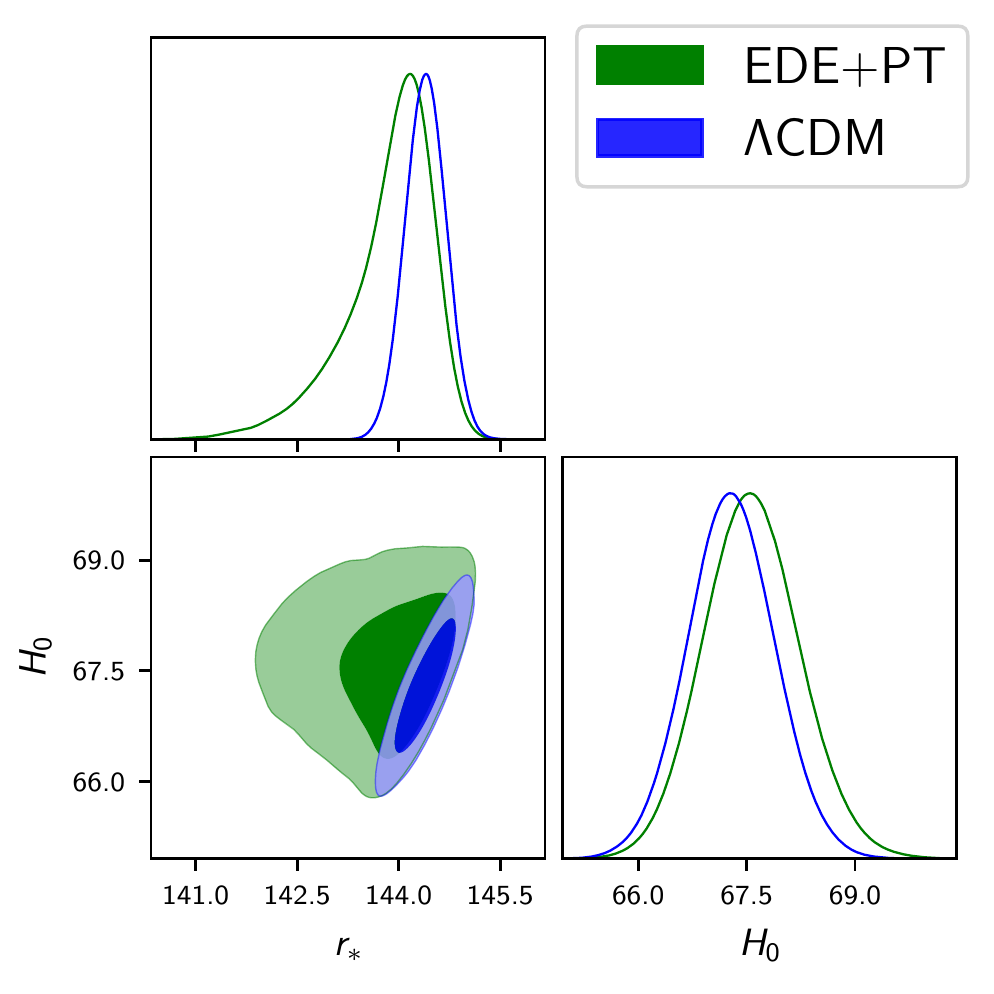}
    \caption{Comparison between the constraints on $H_0$ and $r_*$ obtained from the MCMC analysis using the Planck 2018 data. The blue and green contours show the $\Lambda$CDM and EDE+PT model.}
    \label{fig_mcmc_3412}
\end{figure}

\begin{table}[ht]
    \centering
    \renewcommand{\arraystretch}{1.5}
    \scalebox{0.8}{
    \begin{tabular}{|c|c|c|}
    \hline
    Parameter & $\Lambda$CDM & EDE + PT \\
    \hline
    $100\Omega_{\rm b} h^2$ &$2.236 \left( 2.245 \right) ^{+0.015}_{-0.015}$& $2.241 \left( 2.244 \right)^{+0.017}_{-0.016}$ \\
    \hline
    $\Omega_{\rm c} h^2$ & $0.1202 \left( 0.1196 \right) ^{+0.0013}_{-0.0014}$& $0.1211 \left( 0.1216  \right) ^{+0.0015}_{-0.0021}$ \\
    \hline
    $n_{\rm s}$ & $0.9648 \left( 0.9658 \right) ^{+0.0043}_{-0.0043}$& $0.9645 \left( 0.9648  \right) ^{+0.0049}_{-0.0048}$ \\
    \hline
    $r_s^{*}$ &$144.39 \left( 144.46 \right) ^{+0.30}_{-0.30}$ & $143.85 \left( 143.62 \right) ^{+0.78}_{-0.36}$  \\
    \hline
    $H_0$ & $67.28 \left( 67.53 \right) ^{+0.60}_{-0.59}$& $67.56 \left( 67.69 \right) ^{+0.65}_{-0.66}$ \\
    \hline
    $f_{\rm EDE}$ & - & $0.0145 \left( 0.0153 \right) ^{+0.0034}$\\
    \hline
    $-\mathrm{log}_{10} a_{\rm c}$ & - & $3.97 \left( 3.52 \right) ^{+0.20}_{-0.57}$ \\
    \hline
    $A_{\rm PT}$ & - & $3.41 \left( 4.27 \right) ^{+0.52}_{-1.91}$ \\
    \hline
    $\mathrm{log}_{10}(a_\mathrm{end}/a_\mathrm{c})$ & - & $0.57 \left( 0.86 \right) ^{+0.43}_{-0.47} $  \\
    \hline
    \hline
    $\chi^2$ (CMB) & $2780.17 \left( 2766.11 \right)$ & $2779.93\left( 2765.70 \right)$  \\
    \hline
\end{tabular}
}

    \caption{Mean values and 68\% confidence level for some cosmological parameters in each model obtained from the MCMC analysis using the Planck 2018 data. The best-fit values are shown in the parentheses.}
    \label{tab_params}
\end{table}

\begin{figure}[ht]
    \centering
    \includegraphics[height=15cm]{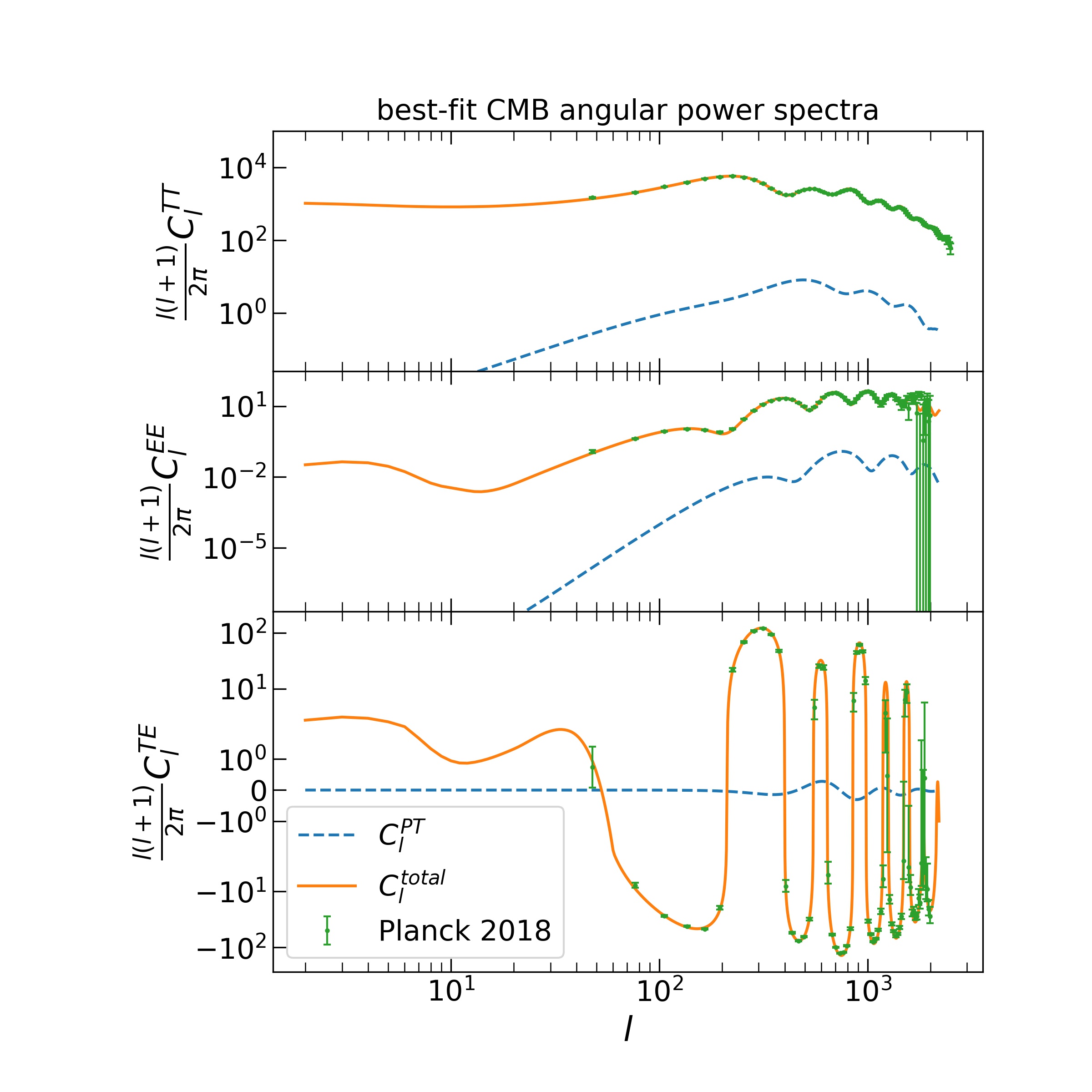}
    \caption{CMB angular power spectra of temperature and E-mode auto-correlations $C_\ell^{TT}$ (top) and $C_\ell^{EE}$ (middle), and temperature E-mode cross-correlation $C_\ell^{TE}$ (bottom) that arise from the EDE PT-mode. The model parameters are fixed to the best-fit values which are obtained from the MCMC analysis using the Planck 2018 data and shown in Table \ref{tab_params}. The TT and EE spectra of the PT mode (blue dashed lines) simply decay in the power-law form at the lower $l$ region. Therefore we only show the ranges of $\frac{\ell (\ell+1)}{2 \pi} C_\ell^{TT} \geq 2.5 \times 10^{-2}$ and $\frac{\ell (\ell+1)}{2 \pi} C_\ell^{EE} \geq 2 \times 10^{-8}$ in those panels.}
    \label{fig_cl_best}
\end{figure}

\begin{figure}[ht]
    \centering
    \includegraphics[height=15cm]{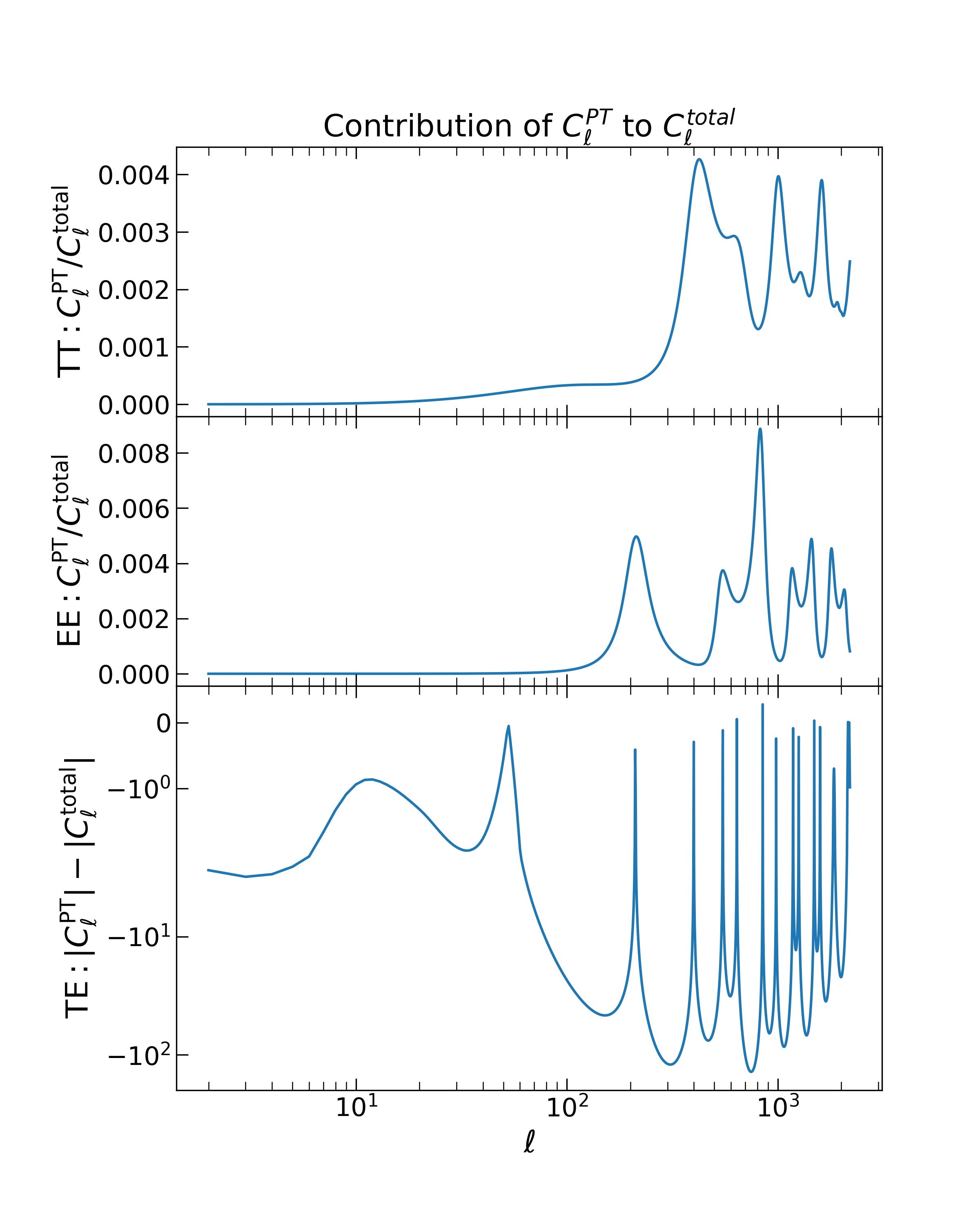}
    \caption{The contribution of $C_\ell^\mathrm{PT}$ to $C_\ell^\mathrm{total}$ obtained from the MCMC analysis using the Planck 2018 data. We plot the fractions of $C_\ell^\mathrm{PT}$ to $C_\ell^\mathrm{total}$ for TT and EE spectra in the upper and middle panels. The contribution of $C_\ell^\mathrm{PT}$ is almost none at large scale. The effects of $C_\ell^\mathrm{PT}$ appear in the high-$\ell$ region. However, the amplitude of $C_\ell^\mathrm{PT}$ is less than 0.5\% of that of $C_\ell^\mathrm{total}$ for the TT spectrum and less than 0.8\% of that of $C_\ell^\mathrm{total}$ for the EE spectrum. For the TE spectrum, the difference between the amplitudes of $C_\ell^\mathrm{PT}$ and $C_\ell^\mathrm{total}$ is shown because the TE spectra cross zero. }
    \label{fig_clvs}
\end{figure}

First, we discuss the constraints on the EDE parameters. We found that smaller values for $f_\mathrm{EDE}, a_c,$ and $A_\mathrm{PT}$ are favored because larger values of them affect the CMB power spectra as discussed in the previous section. In particular $f_\mathrm{EDE}$ is zero-consistent, and we obtained only the upper limits on $f_\mathrm{EDE}$ in Table \ref{tab_params}.
We show the CMB angular spectra induced by the phase transition $C_\ell^{\rm PT}$ and the total spectra $C_\ell^{\rm total}$ in Figure \ref{fig_cl_best}. In this case, we fix the model parameters to the best-fit values in Table \ref{tab_params}.
As shown in Figure \ref{fig_clvs}, the amplitude of $C_\ell^\mathrm{PT}$ is at the sub-percent level of the amplitude of $C_\ell^\mathrm{total}$, so only small contributions of EDE PT-mode are allowed. For the end of the phase transition, we obtained $\mathrm{log}_{10}(a_\mathrm{end}/a_\mathrm{c})=0.57 ^{+0.43}_{-0.47}$. This is the same range as the input prior. As we showed in Figure \ref{fig_claendapt} and discussed in section \ref{sec:results_CMB}, the effect of $a_{\rm{end}}$ on the CMB power spectra is small. Therefore, $a_\mathrm{end}$ is not constrained by the Planck data.

\begin{figure}[ht]
    \centering
    \includegraphics[height=15cm]{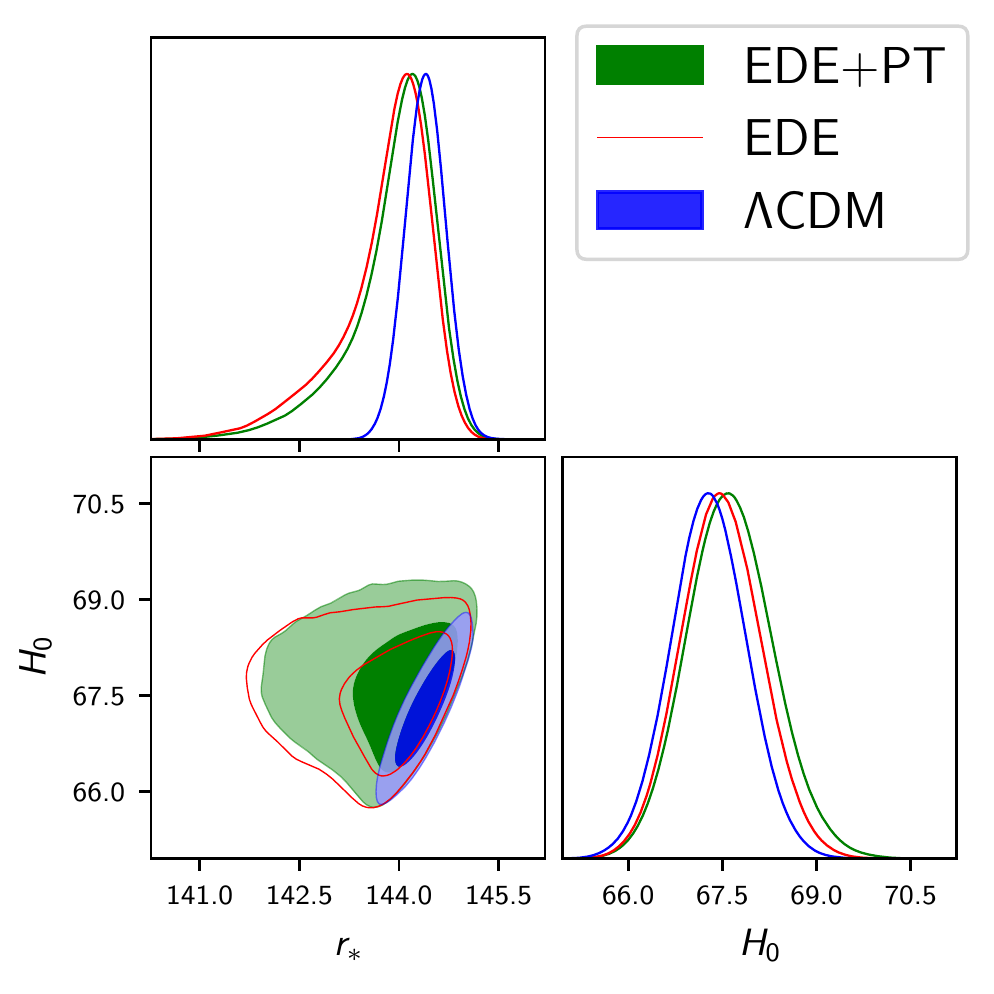}
    \caption{Comparison between the constraints on $H_0$ and $r_\ast$ obtained from the MCMC analysis using the Planck 2018 data. The blue, green and red contours show the standard $\Lambda$CDM model, the EDE model with the PT-mode and the EDE model without the PT-mode.}
    \label{fig_mcmc_3}
\end{figure}

Next, we performed another MCMC analysis to check the effects of PT-mode. In this analysis, we have used the different prior range of $-\mathrm{log}_{10}a_\mathrm{c}$, which is $[3.5, 5.0]$, and fixed $\mathrm{log}_{10}(a_\mathrm{end}/a_\mathrm{c})$ because we have known that this parameter cannot be constrained by previous analysis. According to Figure \ref{fig_mcmc_3}, we also found no significant difference between the constraints in the EDE model with and without the phase transition.
This is because the contribution to the total angular power spectra from PT-mode is at the sub-percent level as shown in Figure \ref{fig_clvs} and this value is so small that the EDE PT-mode does not appear in the CMB spectra.

\begin{table}[ht]
    \centering
    \renewcommand{\arraystretch}{1.5}
    \scalebox{0.8}{
    \begin{tabular}{|c|c|c|}
    \hline
    Parameter & EDE + PT & EDE \\
    \hline
    $100\Omega_{\rm b} h^2$ & $2.243 \left( 2.239 \right) ^{+0.016}_{-0.018} $ & $2.241 \left( 2.230 \right) ^{+0.016}_{-0.016} $ \\
    \hline
    $\Omega_{\rm c} h^2$ & $0.1209 \left( 0.1231 \right) ^{+0.0015}_{-0.0021} $ & $0.1213 \left( 0.1207 \right) ^{+0.0014}_{-0.0022} $ \\
    \hline
    $n_{\rm s}$ & $0.9645 \left( 0.9657 \right) ^{+0.0051}_{-0.0051} $ & $0.9653 \left( 0.9651 \right) ^{+0.0048}_{-0.0048} $ \\
    \hline
    $r_s^{*}$ & $143.91 \left( 143.11 \right) ^{+0.75}_{-0.32} $ & $143.80 \left( 144.23 \right) ^{+0.80}_{-0.32} $ \\
    \hline
    $H_0$ & $67.63 \left( 67.46 \right) ^{+0.65}_{-0.71} $ & $67.46 \left( 67.12 \right) ^{+0.63}_{-0.63} $ \\
    \hline
    $f_{\rm EDE}$ & $0.0146 \left( 0.0245 \right) ^{+0.0032} $ & $0.0176 \left( 0.0075 \right) ^{+0.0038} $ \\
    \hline
    $-\mathrm{log}_{10} a_{\rm c}$ & $4.06 \left( 3.62 \right) ^{+0.18}_{-0.56} $ & $4.15 \left( 4.68 \right) ^{+0.84}_{-0.66} $ \\
    \hline
    $A_{\rm PT}$ & $3.31 \left( 3.51 \right) ^{+0.79}_{-1.45} $ & - \\
    \hline
    \hline
    $\chi^2$ (CMB) & $2780.01 \left( 2766.06 \right)$ & $2781.10 \left( 2768.36 \right)$  \\
    \hline
\end{tabular}
}
    \caption{Mean values and 68\% confidence level for some cosmological parameters in each model obtained from the MCMC analysis using the Planck 2018 data. The best-fit values are shown in parentheses.}
    \label{tab_params2}
\end{table}

As mentioned in Sections \ref{sec:intro} and \ref{sec_2}, an important motivation to introduce the EDE is to resolve the Hubble tension. However, our analysis shows that the constraints on the standard cosmological parameters including $H_0$ do not change significantly in the EDE models and the $\Lambda$CDM model. The least $\chi^2$ values for the EDE models are not improved so much while the number of model parameters is increased by four (``EDE+PT'') and by two (``EDE''). For further discussion on resolving the Hubble tension, we plot the two-dimensional constraint on the sound horizon at the recombination $r_\mathrm{s}^*$ and the Hubble constant $H_0$ in Figure \ref{fig_mcmc_3}. This shows that although the constraints in the EDE models are slightly relaxed in the direction of smaller $r_{\rm s}^{*}$ and larger $H_{\rm 0}$, it is difficult to fully solve the Hubble tension.

\subsubsection{Planck2018+lensing+BAO+supernovae+SH0ES}
\label{MCMC_full}
\begin{figure}[ht]
    \centering
    \includegraphics[height=16cm]{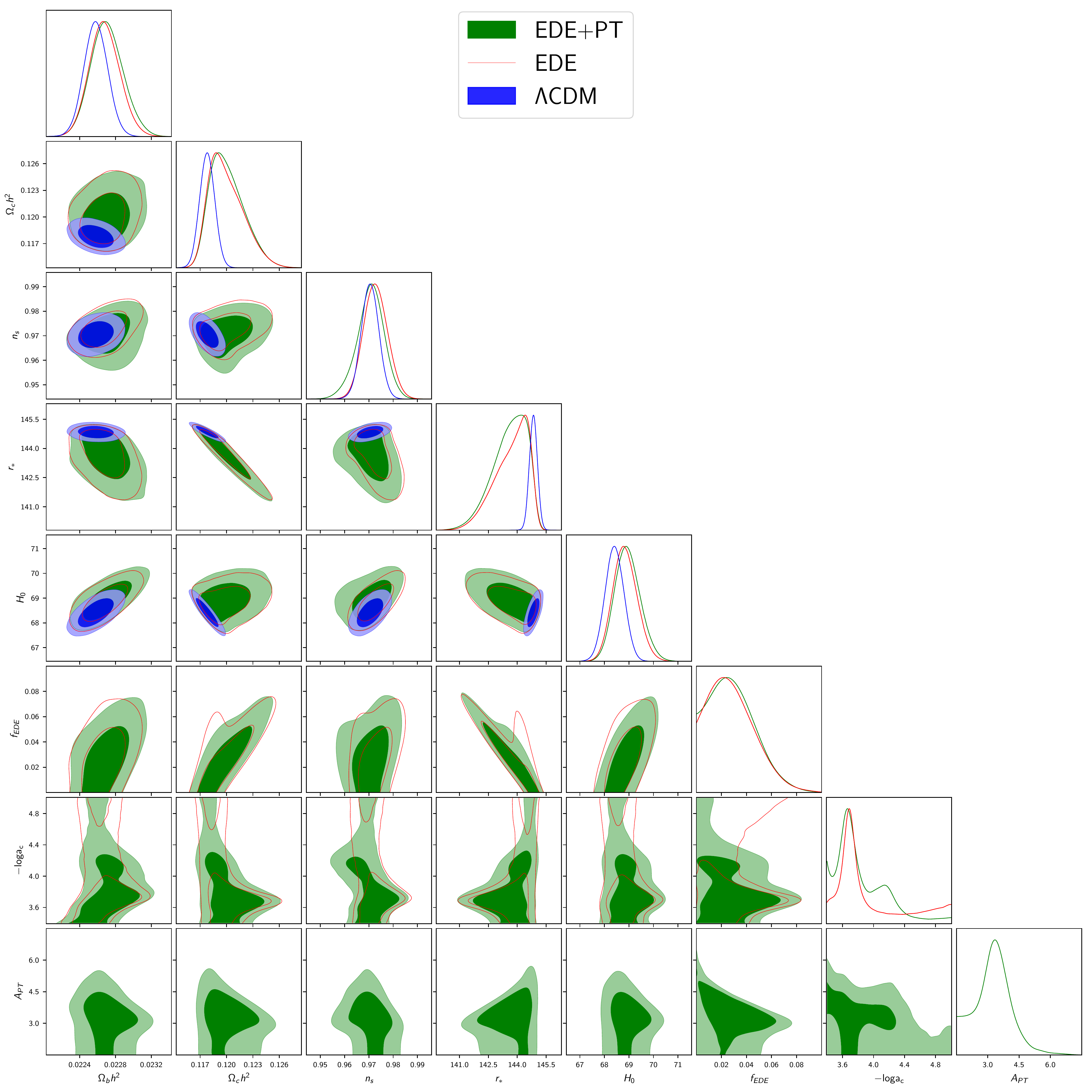}
    \caption{68\% and 95\% confidence level obtained from the MCMC analysis using the Planck 2018+lensing+BAO+SNe+SH0ES data set. The blue, grey and red contour shows the $\Lambda$CDM, EDE+PT and EDE model.}
    \label{fig_mcmc_2}
\end{figure}

We show the results of further analysis by using the Planck2018 CMB data, lensing, BAO, supernovae, and distance ladder data in Figure \ref{fig_mcmc_2}. It can be seen that the trend in Figure \ref{fig_mcmc_3411} is more emphasized in Figure \ref{fig_mcmc_2}.

The best-fitted values and constraints with 68\% confidence level of parameters are shown in Table \ref{tab_params3}. $f_\mathrm{EDE}$ becomes two times lager than that in Table \ref{tab_params}. For the end of the phase transition, we got $\mathrm{log}_{10}(a_\mathrm{end}/a_\mathrm{c})=0.56^{+0.44}_{-0.46}$. Similar to the previous analysis, this is the same as the range of the input prior, so $\mathrm{log}_{10}(a_\mathrm{end}/a_\mathrm{c})$ cannot be constrained. 

The $\chi^2$ values for the SH0ES are shown as $\chi^2 \left(\mathrm{HST}\right)$ in Table \ref{tab_params3}. As shown in Table~\ref{tab_params3}, the best-fitted values of $\chi^2 \left(\mathrm{HST}\right)$ for the EDE and EDE+PT model are 14.37 and 8.55. These values are smaller than that for the $\Lambda$CDM model, $\chi^2 \left( \mathrm{HST} \right) = 19.77$. Therefore our EDE models make the fit to the SH0ES data better. However, we obtained $H_0=68.94^{+0.47}_{-0.57}~\mathrm{km} \, \mathrm{s}^{-1} \mathrm{Mpc}^{-1}$ for the EDE+PT model and $H_0=68.84^{+0.46}_{-0.53}~\mathrm{km} \, \mathrm{s}^{-1} \mathrm{Mpc}^{-1}$ for the EDE model and these values are not large enough to solve the Hubble tension completely when compared to the SH0ES $H_0$ value, $H_0=73.04\pm 1.04~\mathrm{km} \, \mathrm{s}^{-1} \mathrm{Mpc}^{-1}$.

\begin{figure}[ht]
    \centering
    \includegraphics[height=15cm]{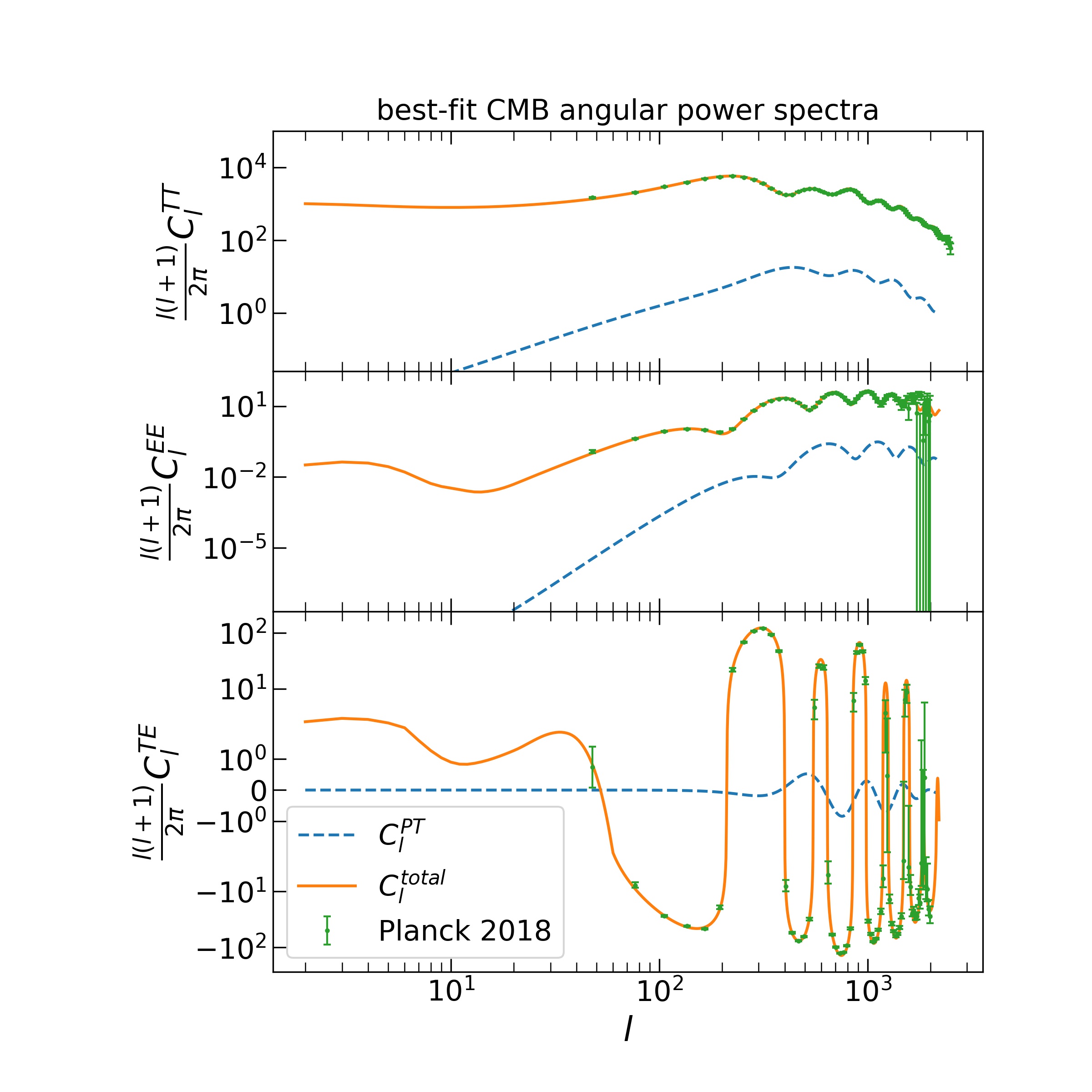}
    \caption{CMB angular power spectra of temperature and E-mode auto-correlations $C_\ell^{TT}$ (top) and $C_\ell^{EE}$ (middle), and temperature E-mode cross-correlation $C_\ell^{TE}$ (bottom) that arise from the EDE PT-mode. The model parameters are fixed to the best-fit values which are obtained by the MCMC analysis using the Planck 2018+lensing+BAO+SNe+SH0ES data set and shown in Table \ref{tab_params3}. The TT and EE spectra of the PT mode (blue dashed lines) simply decay in the power-law form at the lower $l$ region. Therefore we only show the ranges of $\frac{\ell (\ell+1)}{2 \pi} C_\ell^{TT} \geq 2.5 \times 10^{-2}$ and $\frac{\ell (\ell+1)}{2 \pi} C_\ell^{EE} \geq 2 \times 10^{-8}$ in those panels.}
    \label{fig_cl_best2}
\end{figure}

\begin{figure}[ht]
    \centering
    \includegraphics[height=15cm]{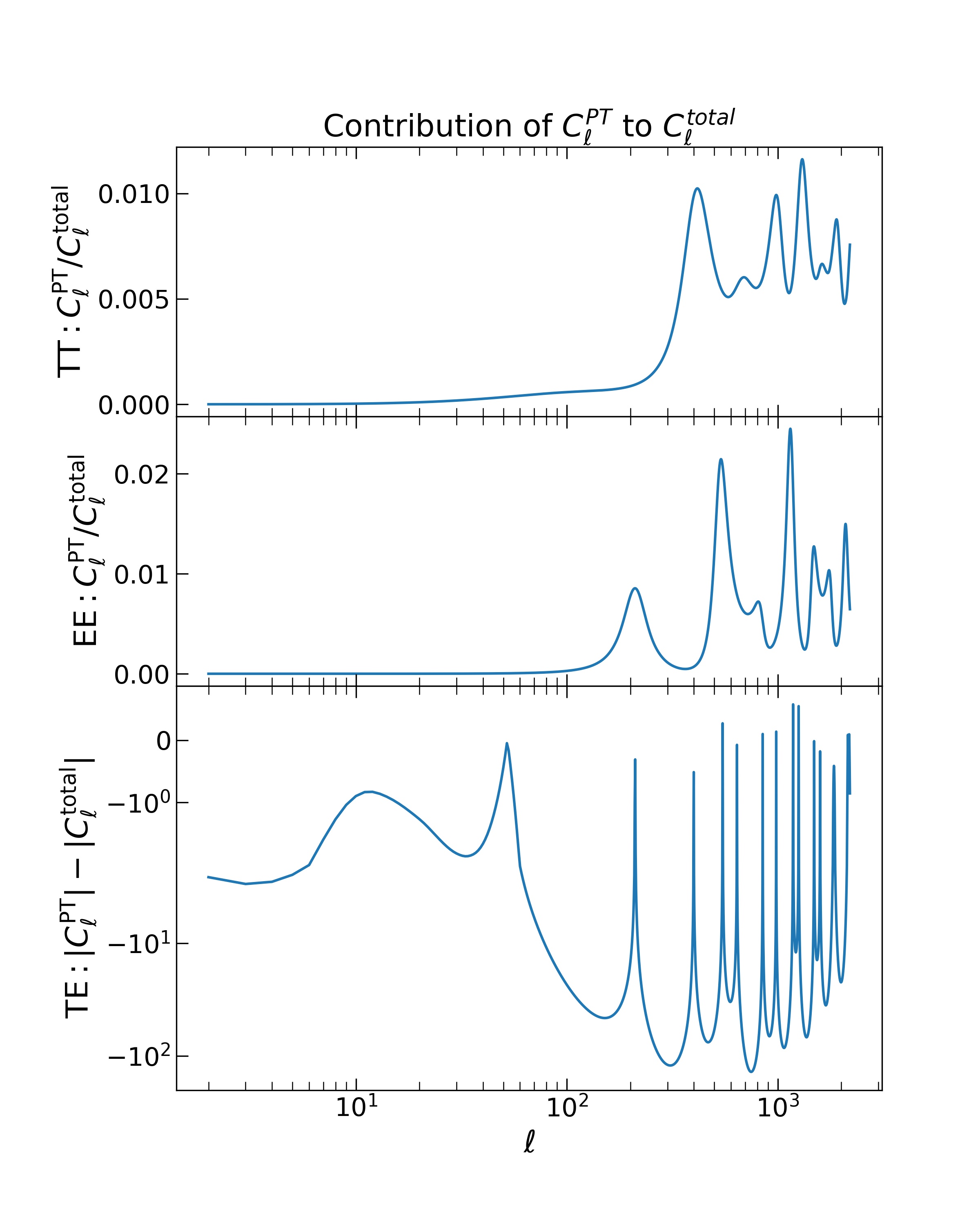}
    \caption{The fractions of $C_\ell^\mathrm{PT}$ to $C_\ell^\mathrm{total}$ for TT and TE spectra, and the difference between $C_\ell^\mathrm{PT}$ and $C_\ell^\mathrm{total}$ for the TE spectrum are shown. The contribution of $C_\ell^\mathrm{PT}$ is almost none at large scales. The effects of $C_\ell^\mathrm{PT}$ appear at higher-$\ell$ multipoles. However the amplitude of $C_\ell^\mathrm{PT}$ is almost 1\% of that of $C_\ell^\mathrm{total}$ at TT spectra and 2\% of that of $C_\ell^\mathrm{total}$ at EE spectra at a maximum.
    Here we used Planck2018+lensing+BAO+SNe+SH0ES data set to obtain model parameters.
    }
    \label{fig_clvs2}
\end{figure}

We plot $C_\ell^\mathrm{total}$ and $C_\ell^\mathrm{PT}$ made by the best-fitted parameters in Figure \ref{fig_cl_best2} and the fraction of $C_\ell^\mathrm{PT}$ to $C_\ell^\mathrm{total}$ in Figure \ref{fig_clvs2}. According to Figure \ref{fig_clvs2}, the maximum contribution of $C_\ell^\mathrm{PT}$ to $C_\ell^\mathrm{total}$ is about 1\% at TT spectra and 2\% at EE spectra. Although this is larger than the results of the analysis of the previous section, this is so small that $C_\ell^\mathrm{PT}$ does not appear in the CMB spectra.

\begin{table}[ht]
    \centering
    \renewcommand{\arraystretch}{1.5}
    \scalebox{0.8}{
    \begin{tabular}{|c|c|c|c|}
    \hline
    Parameter & $\Lambda$CDM & EDE + PT & EDE \\
    \hline
    $100\Omega_{\rm b} h^2$ &$2.258 \left( 2.257  \right) ^{+0.013}_{-0.013}$& $2.270 \left( 2.288 \right)^{+0.017}_{-0.019}$ & $2.268 \left( 2.266 \right) ^{+0.016}_{-0.018}$\\
    \hline
    $\Omega_{\rm c} h^2$ & $0.1178 \left( 0.1178 \right) ^{+0.0008}_{-0.0009}$& $0.1201 \left( 0.1217  \right) ^{+0.0014}_{-0.0024}$ & $0.1200 \left( 0.120 \right) ^{+0.0014}_{-0.0024}$\\
    \hline
    $n_s$ &$0.9705 \left( 0.9717 \right) ^{+0.0036}_{-0.0036}$ &$0.9707 \left( 0.9777 \right) ^{+0.0059}_{-0.0050}$ &$0.9728 \left( 0.9756 \right) ^{+0.0047}_{-0.0051}$ \\
    \hline
    $r_s^{*}$ &$144.84 \left( 144.84 \right) ^{+0.21}_{-0.21}$ & $143.61 \left( 142.49 \right) ^{+1.15}_{-0.58}$ & $143.69 \left( 142.84 \right) ^{+1.14}_{-0.53}$ \\
    \hline
    $H_0$ & $68.41 \left( 68.42 \right) ^{+0.38}_{-0.38}$& $68.94 \left( 70.00 \right) ^{+0.47}_{-0.57}$ & $68.84 \left( 69.10 \right) ^{+0.46}_{-0.53}$\\
    \hline
    $f_{\rm EDE}$ & - & $0.029 \left( 0.054 \right) ^{+0.007}$& $0.029 \left( 0.042 \right) ^{+0.011}$\\
    \hline
    $-\mathrm{log}_{10} a_{\rm c}$ & - & $3.84 \left( 3.65 \right) ^{+0.08}_{-0.44}$ & $3.93 \left( 3.66 \right) ^{-0.02}_{-0.53}$\\
    \hline
    $A_{\rm PT}$ & - & $3.26 \left( 3.44 \right) ^{+0.92}_{-0.79}$ & - \\
    \hline
    $\mathrm{log}_{10}(a_\mathrm{end}/a_\mathrm{c})$ & - & $0.56 \left( 0.34 \right) ^{+0.44}_{-0.46} $ & - \\
    \hline
    \hline
    $\chi^2$ (CMB+lensing) & $2792.27 \left( 2780.80 \right)$ & $2792.66 \left( 2780.57 \right)$ & $2794.23 \left( 2779.09 \right)$ \\
    \hline
    $\chi^2$ (HST) & $19.95 \left( 19.77 \right)$ & $15.77 \left(8.55  \right)$ & $16.51 \left( 14.37 \right)$ \\
    \hline
    $\chi^2$ (BAO) & $8.29 \left( 7.95 \right)$ & $8.31 \left( 9.01 \right)$ & $8.27 \left( 7.92 \right)$ \\
    \hline
    $\chi^2$ (SN) & $1034.80 \left( 1034.75 \right)$ & $1034.81 \left( 1034.75 \right)$ & $1034.82 \left( 1034.79 \right)$ \\
    \hline
    \hline
    $\chi^2$ (total) & $3855.31 \left( 3843.27 \right)$ & $3851.55 \left( 3832.24 \right)$ & $3853.83 \left( 3836.17 \right)$ \\
    \hline
\end{tabular}
}
    \caption{The results obtained from the MCMC analysis using the Planck 2018 + lensing + BAO + SNe + SH0ES data set. Mean values and 68\% confidence level for some cosmological parameters in each model. The best-fit values are shown in parentheses. $\chi^2\mathrm{(HST)}$ is the $\chi^2$ value for the SH0ES.}
    \label{tab_params3}
\end{table}

\begin{figure}[ht]
    \centering
    \includegraphics[height=15cm]{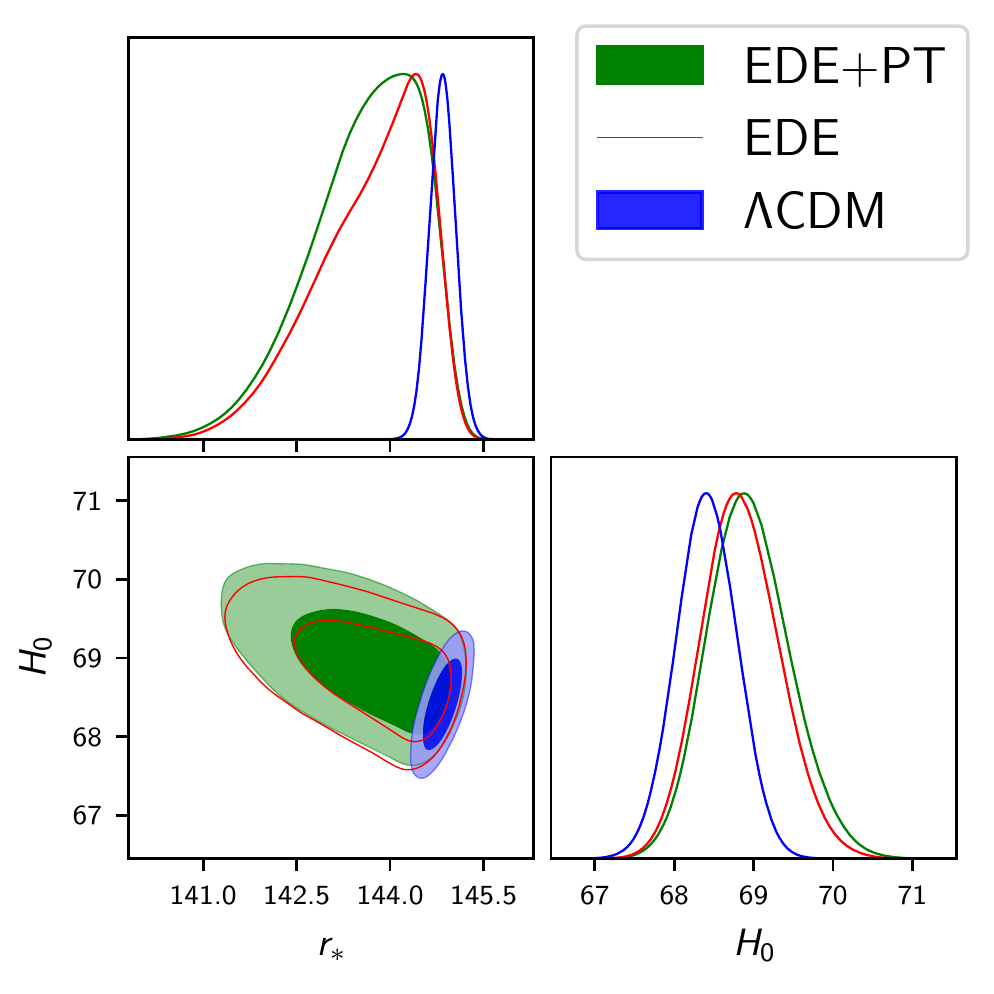}
    \caption{Comparison between the constraints on $H_0$ and $r_\ast$ obtained from the Planck2018+lensing+BAO+SNe+SH0ES data set. The blue, green, and red contours show the standard $\Lambda$CDM model, the EDE model with the PT-mode and the EDE model without the PT-mode.}
    \label{fig_mcmc_4}
\end{figure}

\section{Summary}
\label{sec:4}
In this paper, we phenomenologically investigated the effects on CMB anisotropies at background and perturbation levels of the EDE model with phase transition. We then provided observational constraints from the Planck data. The phase transition of the EDE should occur at different epochs for different horizon patches, producing the ``PT-mode'' cosmological perturbations, which are independent of adiabatic ones predicted by inflationary mechanisms. We calculated the time evolution of the PT-mode perturbations and the resultant CMB angular power spectrum for the first time, and constrained the parameters of $\Lambda$CDM cosmology and the EDE model.

We modified the Boltzmann code CAMB by adding source terms into the perturbed equations for the EDE fluid to introduce the PT-mode. Here we used a power-law type initial power spectrum to determine the PT-mode source terms. For simplicity, we considered a white noise-like initial power spectrum for the PT-mode, and we set $n_\mathrm{PT}=4$. We performed the MCMC analysis by using Planck 2018 data and other data from Planck 2018 CMB lensing, BAO, supernovae, and SH0ES, based on our modified CAMB code. As a result, the amplitude of the PT-mode perturbation was limited to less than $10^{-4}$ at 95\% confidence level. This is so small that the effect does not appear in the CMB angular spectra. Thus, we conclude that the simple PT-mode EDE model cannot solve the Hubble tension. Furthermore, it is difficult to constrain the end of the phase transition in all our analyses.

Finally, for axion-like EDE models, it is argued that performing the analysis beyond the fluid approximation is important to get the correct results \cite{2018PhRvD..98h3525P, 2020PhRvD.101b3501C}. 
However, in our analysis, there does not appear to be a significant difference between the results under the fluid approximation and those obtained by solving the Klein-Gordon equation.
Moreover, we consider a more general EDE model and do not specify the EDE model to an axion-like model. For the above reasons, we assume the fluid implementation of the model. It is worth investigating cosmological constraints beyond the fluid approximation. However, because it is beyond the scope of this paper, we put them as future works.

\section{Acknowledgement}
We would like to thank L. Herold and S. Yokoyama for useful discussions. This work is supported in part by JSPS Overseas Research Fellowship (TM), and the JSPS grant numbers 18K03616, 21H04467 and JST AIP Acceleration Research Grant JP20317829 and JST FOREST Program JPMJFR20352935 (KI). We thank the anonymous referee for the helpful comments and advice.

\clearpage 
\bibliography{EDE}

\end{document}